\documentclass[11pt,pre,byrevtex,superscriptaddress,bibnotes,showkeys]{revtex4}
\usepackage{epsfig}
\usepackage{times}

\newtheorem{defn}{Definition}[section]

\newtheorem{thm}[defn]{Theorem}
\newtheorem{prop}[defn]{Proposition}
\newtheorem{lemma}[defn]{Lemma}

\newtheorem{remark}[defn]{Remark}
\newtheorem{example}[defn]{Example}

\newcommand{\ve}{\varepsilon}
\newcommand{\R}{I\!\!R}

\newcommand{\E}{\mathcal{E}}
\newcommand{\N}{{I\!\!N}}
\newcommand{\F}{\mathcal{F}}
\newcommand{\noi}{\noindent}
\newcommand{\vsp}{\vspace{.2in}}
\newcommand{\hsp}{\hspace{.2in}}
\newcommand{\goto}{\rightarrow}

\newcommand{\bea}{\begin{eqnarray}}
\newcommand{\eea}{\end{eqnarray}}
\newcommand{\beas}{\begin{eqnarray*}}
\newcommand{\eeas}{\end{eqnarray*}}
\newcommand{\be}{\begin{equation}}
\newcommand{\ee}{\end{equation}}
\renewcommand{\P}{\mathcal P}

\newcommand{\per}{\hspace{-.072in}{\bf .  }}

\newcommand{\ebk}{{\cal E}_{\beta,K}}
\newcommand{\tebk}{\tilde{{\cal E}}_{\beta,K}}
\newcommand{\euk}{{\cal E}^{u,K}}

\newcommand{\tz}{\tilde{z}}
\newcommand{\tw}{\tilde{w}}
\newcommand{\kconeb}{K_c^{(2)}(\beta)}
\newcommand{\kctwob}{K_c^{(1)}(\beta)}
\newcommand{\ktc}{K_{\mbox{{\scriptsize \rm tri}}}^{\mbox{{\scriptsize \rm canon}}}}
\newcommand{\ktm}{K_{\mbox{{\scriptsize \rm tri}}}^{\mbox{{\scriptsize \rm micro}}}}
\newcommand{\skp}{\vspace{\baselineskip}}

\newcommand{\brmk}{\begin{remark}\per\begin{em}}
\newcommand{\ermk}{\end{em}\end{remark}}

\newcommand{\bexa}{\begin{example}\per\begin{em}}
\newcommand{\eexa}{\end{em}\end{example}}
\newcommand{\ink}{\rule{.5\baselineskip}{.55\baselineskip}}

\newcommand{\beginsec}{\setcounter{equation}{0}}

\begin{document}

\title{Analysis of Phase Transitions in the Mean-Field
Blume-Emery-Griffiths Model}
\author{R.S. Ellis}
\email{rsellis@math.umass.edu}
\affiliation{Department of Mathematics and Statistics, University of
Massachusetts, Amherst, MA. USA 01003}
\author{P. Otto}
\email{potto@gettysburg.edu}
\affiliation{Department of Mathematics, Gettysburg College,
Gettysburg, PA, USA 17325}
\author{H. Touchette}
\email{htouchet@alum.mit.edu}
\affiliation{School of Mathematical Sciences, Queen Mary, University of London,
London, UK E1 4NS}

\begin{abstract}
In this paper we give a complete analysis of the phase transitions in the
mean-field Blume-Emery-Griffiths lattice-spin model with respect to the canonical ensemble,
showing both a second-order, continuous phase transition and a first-order, discontinuous
phase transition for appropriate values of the thermodynamic
parameters that define the model.
These phase transitions are analyzed both in terms of the empirical measure
and the spin per site by studying bifurcation phenomena of the corresponding sets
of canonical equilibrium macrostates, which are defined via large deviation principles.
Analogous phase transitions with respect to the microcanonical
ensemble are also studied via a combination of rigorous analysis and numerical
calculations.  Finally, probabilistic limit theorems for appropriately scaled values of the total
spin are proved with respect to the canonical ensemble.  These limit theorems include both
central-limit-type theorems when the thermodynamic parameters are not equal
to critical values and non-central-limit-type theorems when these parameters equal
critical values.
\end{abstract}



\maketitle

\section{Introduction}
\beginsec

The Blume-Emery-Griffiths (BEG) model \cite{BEG} is an important lattice-spin model in statistical mechanics.  It is one of the few and certainly one of the simplest models known to exhibit, in the mean-field approximation, both a continuous, second-order phase transition and a discontinuous, first-order phase transition.  Because of this property, the model has been studied extensively as a model of many diverse systems including He$^3$-He$^4$ mixtures --- the system for which Blume, Emery, and Griffiths first devised their model \cite{BEG} --- as well as solid-liquid-gas systems \cite{LS,SL,SL2}, microemulsions \cite{SS}, semiconductor alloys \cite{ND}, and electronic conduction models \cite{KEL}.  On a more theoretical level, the BEG model has also played an important role in the development of the renormalization-group theory of phase transitions of the Potts model; see \cite{HB, NBRS} for details and references.

As a long-range model with a simple description but a relatively complicated
phase transition structure, the BEG model continues to be of interest in modern
statistical mechanical studies.  Our motivation for revisiting this model was initiated by a recent observation in 
\cite{BMR,BMR2} that the mean-field version of the BEG model has nonequivalent microcanonical and canonical ensembles, in the sense that it exhibits microcanonical equilibrium properties having no equivalent within the canonical ensemble.
This observation has been verified in \cite{ETT} by numerical calculations both at the thermodynamic level, as in 
\cite{BMR,BMR2}, and at the level of equilibrium macrostates.  In response to these earlier works, in this paper we address the phase transition behavior of the model by giving separate analyses of 
the structure of the sets of equilibrium macrostates for each of the two ensembles.  Not only are our results consistent with the findings in \cite{BMR,BMR2,ETT}, but also we rigorously prove for the first time a number of results that significantly generalize those found in these papers, where they were derived nonrigorously.  For the canonical ensemble, full proofs of the structure of the
set of equilibrium macrostates are provided.  For the microcanonical ensemble, full proofs could not be attained.  However, using numerical methods and following an analogous technique used in the canonical case, we also analyze the structure of the set of microcanonical equilibrium macrostates.

The BEG model is a spin-1 model defined on the set $\{1,2,...,n\}$.  The
spin at site $j \in \{1,2,...,n\}$ is denoted by $\omega_j$, a
quantity taking values in $\Lambda \doteq \{-1,0,1\}$.  The Hamiltonian
for the BEG model is defined by
\[
H_{n,K}(\omega) \doteq \sum_{j=1}^n \omega_j^2 - \frac{K}{n}
 \left(\sum_{j=1}^n \omega_j \right)^2,
\]
where $K>0$ is given and $\omega = (\omega_1,...,\omega_n) \in \Lambda^n$.  The
energy per particle is defined by
\be
\label{eqn:energyperparticle}
h_{n,K}(\omega) \doteq \frac{1}{n} H_{n,K}(\omega) =
\frac{\sum_{j=1}^n \omega_j^2}{n} - K \left(\frac{\sum_{j=1}^n
\omega_j}{n} \right)^2.  \ee

In order to analyze the phase transition behavior of the model,
we first introduce the sets of equilibrium macrostates for the canonical 
ensemble and the microcanonical ensemble.  
As we will see, the canonical equilibrium
macrostates solve a two-dimensional, unconstrained minimization problem while the
microcanonical equilibrium macrostates solve a dual, one-dimensional, constrained
minimization problem.  
The definitions of these sets follow from large deviation principles
derived for general models in \cite{EHT1}.  In the particular case of
the BEG model they are consequences of the fact that the BEG-Hamiltonian
can be written as a function of the 
empirical measures of the spin random variables and that 
according to Sanov's Theorem the large
deviation behavior of these empirical measures is governed by
the relative entropy.

We use two innovations to analyze
the structure of the set of canonical equilibrium macrostates.  
The first is to 
reduce to a one-dimensional problem the two-dimensional minimization problem that characterizes 
these macrostates.  This is carried out by absorbing the noninteracting
component of the energy per particle function into the prior measure, which is a product measure
on configuration space.  This manipulation allows us to express
 the canonical ensemble in terms of the empirical
means, or spin per site, of the spin random variables.  
Doing so reduces the analysis of BEG model to the analysis of a Curie-Weiss-type model \cite{Ellis}
with single-site measures depending on $\beta$.  

The analysis of the set of canonical equilibrium macrostates is further simplified by a second innovation.  
Because the thermodynamic parameter that defines the canonical ensemble is the inverse temperature
$\beta$, a phase transition with respect to this ensemble is defined by fixing the Hamiltonian-parameter $K$ 
and varying 
$\beta$.   Our analysis of the set of canonical equilibrium macrostates is based on a 
much more efficient approach that 
fixes $\beta$ and varies $K$.  
Proceeding in this way allows us to solve rigorously and in complete detail
the reduced one-dimensional problem characterizing the equilibrium macrostates.  We then 
extrapolate these results obtained by fixing $\beta$ and varying $K$ to physically relevant results that hold 
for fixed $K$ and varying $\beta$.  These include a second-order, continuous phase transition and a first-order, discontinuous phase transition for different ranges of $K$.

For the microcanonical ensemble, we use a technique employed in \cite{BMR} that absorbs the constraint into the minimizing function.  This
step allows us to reduce the constrained minimization
problem defining the microcanonical equilibrium macrostates
to a one-dimensional, unconstrained minimization problem.
Rigorous analysis of the reduced problem being
limited, we rely mostly on numerical computations to complete our analysis of the set of equilibrium macrostates. 
Because the thermodynamic parameter defining the microcanonical ensemble is the energy per particle $u$, a phase transition with respect to this ensemble is defined by fixing $K$ and varying 
$u$.  By analogy with the canonical case, 
our numerical analysis of the set of microcanonical equilibrium macrostates is based on a much more efficient 
approach that fixes $u$ and varies $K$.
The analysis with respect to $K$ rather than $u$ allows us to solve in some detail
the reduced one-dimensional problem characterizing the equilibrium macrostates.  We then 
extrapolate these results obtained by fixing $u$ and varying $K$ to physically relevant results that hold 
for fixed $K$ and varying $u$.  As in the case of the canonical ensemble, these include a second-order, continuous phase transition and a first-order, discontinuous phase transition for different ranges of $K$.

The contributions of this paper include a rigorous global analysis of the first-order phase transition in the canonical ensemble.  Blume, Emery, and Griffiths did a local analysis of the spin per site to show that their model exhibits a second-order phase transition for a range of values of $K$ and that at a certain value of $K$ a tricritical point appears \cite{BEG}.  This tricritical point has the property that for all smaller values of $K$, we are dealing with a first-order phase transition.  Mathematically, the tricritical point marks the beginning of the failure of the local analysis; beyond this point one has to resort to a global analysis of the spin per site.  While the first-order phase transition has been studied numerically by several authors, the present paper gives the first rigorous global analysis.  

Another contribution is that we analyze the phase transition for the canonical ensemble both in terms
of the spin per site and the empirical measure.  While all previous studies 
of the BEG model except for \cite{ETT} focused only on the spin per site, the analysis in terms
of the empirical measure is the natural context for understanding 
equivalence and nonequivalence of ensembles \cite{ETT}.

A main consequence of our analysis is that 
the tricritical point --- the critical value of the Hamiltonian
parameter $K$ at which the model changes its phase transition behavior
from second-order to first-order --- differs in the two ensembles.
Specifically, the tricritical point
is smaller in the microcanonical ensemble than in the canonical ensemble.
Therefore, there exists a range of values of $K$ such that the BEG
model with respect to the canonical ensemble exhibits a first-order phase
transition while with respect to the microcanonical ensemble the
model exhibits a second-order phase transition.  As we discuss in Section 5, 
these results are consistent with the observation, shown numerically in 
\cite{ETT}, that there
exists a subset of the microcanonical equilibrium macrostates
that are not realized canonically.  This observation implies that the two ensembles are
nonequivalent at the level of equilibrium macrostates.  

A final contribution of this paper is to present probabilistic limit theorems  
for appropriately scaled partial sums $S_n \doteq \sum_{j=1}^n \omega_j$ with respect to the canonical ensemble. 
These limits, which follow from our work in Section 3 and known limit theorems for the Curie-Weiss model 
derived in \cite{EN,ENR}, include conditioned limit theorems when there are 
multiple phases.  In most cases the limits involve the central-limit-type
scaling $n^{1/2}$ and convergence in distribution of $S_n/n^{1/2}$ to a 
normal random variable.
They also include the following two nonclassical cases, which hold for appropriate critical values of the parameters defining the canonical ensemble: 
\[
S_n/n^{3/4} \stackrel{{\cal D}}{\longrightarrow} X, \mbox{where } P\{X \in dx\} =  \mbox{const} \cdot 
\exp[-\mbox{const} \cdot x^4] \, dx
\]
and
\[
S_n/n^{5/6} \stackrel{{\cal D}}{\longrightarrow} X, \mbox{where } P\{X \in dx\} =  \mbox{const} \cdot 
\exp[-\mbox{const} \cdot x^6] \, dx.
\]
As in the case of more complicated models such as the Ising model, 
these nonclassical theorems signal the onset of
a phase transition in the BEG model \cite[Sect.\ V.8]{Ellis}.
They are analogues of a result for the much simpler Curie-Weiss model
\cite[Thm.\ V.9.5]{Ellis}.  

The outline of the paper is as follows.  In Section 2, 
following the general procedure described in \cite{EHT1}, we 
define the canonical ensemble, the microcanonical ensemble, and the
corresponding sets of equilibrium macrostates.  In Section 3, we outline our analysis of the structure of the set
of canonical equilibrium macrostates.   Because the proofs involve many technicalities,
for ease of exposition we rely in this section on graphical 
arguments which, though not rigorous, convincingly
motivate the truth of our assertions.  The interested reader
is referred to \cite{O} for full details and complete rigor.  In Section 4, we present new theoretical insights into, and numerical results concerning, the structure of the set of microcanonical equilibrium macrostates.
In Section 5, we discuss the implications of the results in the two previous sections concerning the 
nature of the phase transitions in the BEG model, which in turn is 
related to the phenomenon of ensemble nonequivalence
at the level of equilibrium macrostates.  Section 6 is devoted to probabilistic
limit theorems for appropriately scaled sums $S_n$.


\section{Sets of Equilibrium Macrostates for the Two Ensembles}
\beginsec

The canonical and microcanonical ensembles are defined in terms of
probability measures on a sequence of
probability spaces $(\Lambda^n, \F_n)$.  The configuration spaces
$\Lambda^n$ consist of microstates
$\omega = (\omega_1,...,\omega_n)$ with each $\omega_j \in \Lambda
\doteq
\{-1,0,1\}$, and $\F_n$ is the $\sigma$-field consisting of all subsets
of $\Lambda^n$.  We also introduce the $n$-fold product measure $P_n$ 
on $\Omega_n$ with
identical one-dimensional marginals 
$\rho \doteq \frac{1}{3}(\delta_{-1} + \delta_0 +
\delta_1)$.  

In terms of the energy per particle $h_{n,K}$ defined
in (\ref{eqn:energyperparticle}), for each $n \in \N$, $\beta >
0$, and $K > 0$ the partition function is defined by
\[
Z_n(\beta,K) \doteq \int_{\Lambda^n} \exp\!\left[ -n \beta
h_{n,K} \right] dP_n.
\]
For sets $B \in \F_n$, the
canonical ensemble for the BEG model is the probability measure
\be
\label{eqn:canon}
P_{n,\beta,K}(B) \doteq \frac{1}{Z_n(\beta,K)} \, \cdot \int_B \exp\!
\left[ -n \beta h_{n,K}\right] dP_n.  
\ee
For
$u \in \R$, $r > 0$, $K > 0$ and sets $B \in \F_n$, the microcanonical
ensemble is the conditional probability measure
\bea
\label{eqn:microensemble}
P_n^{u,r,K} (B) & \doteq & P_n \{ B \ | \ h_{n,K} \in [u-r, u+r] \} \\
& = & \frac{P_n \{ B \cap \{h_{n,K} \in [u-r, u+r] \}\}}{P_n \{h_{n,K}
\in
[u-r, u+r] \}}. \nonumber
\eea
As we point out after (\ref{eqn:microentropy}), for appropriate
values of $u$ and all sufficiently large $n$ the denominator is positive
and thus $P_n^{u,r,K}$ is well defined.

The key to our analysis of the BEG model is to express both the
canonical and the microcanonical ensembles in terms of the
empirical measure $L_n$ defined for $\omega \in \Lambda^n$ by
\[
\label{eqn:empmeasure}
L_n = L_n(\omega, \cdot) \doteq \frac{1}{n} \sum_{j=1}^n
\delta_{\omega_j} (\cdot).  
\]
$L_n$ takes values in $\mathcal{P} =
\mathcal{P}(\Lambda)$, the set of probability measures on $\Lambda$.
We rewrite $h_{n,K}$ as 
\beas
\label{eqn:empiricalmeasureenergy}
h_{n,K}(\omega) & \doteq & \frac{\sum_{j=1}^n \omega_j^2}{n} - K
\left(\frac{\sum_{j=1}^n \omega_j}{n} \right)^2  \\ & = &
\int_\Lambda y^2 L_n(\omega,dy) - K \left(\int_\Lambda y
L_n(\omega,dy) \right)^2.  \nonumber
\eeas 
For $\mu \in \mathcal{P}$ define 
\bea
\label{eqn:interactfcn}
f_K(\mu) & \doteq & \int_{\Lambda} y^2 \mu(dy) -
K\left(\int_{\Lambda} y \mu(dy) \right)^2 \\
& = & (\mu_1 + \mu_{-1}) - K(\mu_1 - \mu_{-1})^2.
\nonumber
\eea
The range of this function 
is the closed interval $[(1-K)\wedge 0, 1]$. 
In terms of $f_K$ we express $h_{n,K}$ in the form
\[
\label{eqn:energyasinteraction}
h_{n,K}(\omega) = f_K (L_n(\omega)).
\]

We appeal to the theory of large deviations to define the sets of canonical
equilibrium macrostates and microcanonical equilibrium macrostates.
Since any $\mu \in \mathcal{P}$ has the form $\sum_{i=-1}^1 \mu_i \delta_i$,
where $\mu_i \geq 0$ and $\sum_{i=-1}^1\mu_i= 1$, $\mathcal{P}$
can be identified with the set of probability vectors in $\R^3$.
We topologize $\mathcal{P}$ with the relative topology that this set
inherits as a subset of $\R^3$.  
The relative entropy of
$\mu \in \mathcal{P}$ with respect to $\rho$ is defined by
\[
\label{eqn:relentropy}
R(\mu|\rho) \doteq \sum_{i=-1}^1 \mu_i \log (3\mu_i).
\] 
Sanov's Theorem states that with respect to the
product measures $P_n$, the empirical measures
$L_n$ satisfy the large deviation principle (LDP) with rate function
$R(\cdot|\rho)$ \cite[Thm.\ VIII.2.1]{Ellis}.  That is, for 
any closed subset $F$ of $\mathcal{P}$ we have the large deviation
upper bound
\[
\limsup_{n \goto \infty} \frac{1}{n} \log P_n\{L_n \in F\} \leq
- \inf_{\mu \in F} R(\mu|\rho),
\]
and for any open subset $G$ of $\mathcal{P}$ we have the large deviation
lower bound
\[
\limsup_{n \goto \infty} \frac{1}{n} \log P_n\{L_n \in G\} \geq
- \inf_{\mu \in G} R(\mu|\rho).
\]

From the LDP for the $P_n$-distributions of
$L_n$, we can derive the LDPs of $L_n$ with
respect to
the two ensembles $P_{n,\beta,K}$ and $P_n^{u,r,K}$.  
In order to state these LDPs, we introduce
two basic thermodynamic functions, one associated with each ensemble.
For $\beta > 0$ and $K > 0$, the basic thermodynamic function for the canonical
ensemble is the canonical free energy 
\[
\label{eqn:freeenergy}
\varphi_K (\beta) \doteq - \lim_{n \goto \infty} \frac{1}{n} \log
Z_n(\beta,K).  
\] 
It follows from Theorem 2.4(a) in \cite{EHT1} that this limit exists
for all $\beta > 0$ and $K > 0$ and
is given by
\[
\label{eqn:freeenergy2}
\varphi_K (\beta) = \inf_{\mu \in
\mathcal{P}} \{R (\mu |\rho) + \beta f_K(\mu)\}.
\]
For the microcanonical ensemble, the basic
thermodynamic function is the microcanonical entropy 
\be
\label{eqn:microentropy}
s_K(u) \doteq -\inf \{ R(\mu | \rho) : \mu \in \mathcal{P}, f_K (\mu)
= u \}.
\ee 
Since $R(\mu | \rho) \geq 0$ for all $\mu$, $s_K(u) \in [-\infty,0]$ for all $u$.
We define $\mbox{dom} \, s$ to be the set of $u \in \R$ for which 
$s_K(u) > -\infty$.  Clearly, $\mbox{dom} \, s_K$ coincides with the
range of $f_K$ on ${\cal P}$, which equals the closed interval
$[(1-K)\wedge 0,1]$.  
For $u \in \mbox{dom} \, s_K$ and all sufficiently large $n$
the denominator in the second line of (\ref{eqn:microensemble})
is positive and thus the microcanonical ensemble $P_n^{u,r,K}$ is well defined 
\cite[Prop.\ 3.1]{EHT1}.

The LDPs for $L_n$ with respect to the two ensembles are given in the
next theorem.  They are consequences of Theorems 2.4 and 3.2 in 
\cite{EHT1}.

\begin{thm} \per  {\em (a)} With respect to the canonical ensemble 
$P_{n,\beta,K}$, the empirical measures $L_n$ satisfy
the {\em LDP} with rate function
\be
\label{eqn:canonrate}
I_{\beta,K}(\mu) = R(\mu |\rho) + \beta f_K(\mu) -
\varphi_K(\beta).
\ee

{\em (b)}  With respect to the microcanonical ensemble $P_n^{u,r,K}$,
the empirical measures $L_n$ satisfy
the {\em LDP}, in the double limit $n \goto \infty$ and $r \goto 0$, with rate function
\be
\label{eqn:microrate}
I^{u,K} (\mu) \doteq \left\{ \begin{array}{ll} R(\mu | \rho) +
s_K (u) &
\mbox{if $f_K (\mu) = u$} \\ \infty & \mbox{otherwise}.
\end{array}
\right.  
\ee
\end{thm}

For $\mu \in \P$ and $\ve > 0$ we denote by $B(\mu,\ve)$
the closed ball in $\P$ with center $\mu$ and radius $\ve$.  
If $I_\beta(\mu) > 0$, then for all sufficiently small $\ve > 0$,
$\inf_{\nu \in B(\mu,\ve)} I_\beta(\mu) >0$.
Hence, by the large deviation upper bound for $L_n$ with respect to 
the canonical ensemble, for all $\mu
\in \mathcal{P}$ satisfying $I_{\beta}(\mu) > 0$, all
sufficiently small $\ve > 0$, and all sufficiently large $n$
\[
P_{n,\beta,K} \{ L_n \in B(\mu, \ve) \} \leq \exp\!\left[-n \textstyle\!\left(\inf_{\nu \in B(\mu,\ve)}
I_\beta(\nu)\right)/2\right] \!,
\] 
which converges to 0 exponentially fast.
Consequently, the most probable
macrostates $\nu$ solve $I_{\beta,K}(\nu) = 0$.  It is 
therefore natural to
define the set of canonical equilibrium macrostates
to be 
\bea
\label{eqn:canonmacrostates}
\mathcal{E}_{\beta,K} & \doteq &
\{\nu \in \mathcal{P} : I_{\beta,K}(\nu) = 0 \} \\ & = & \{\nu \in
\mathcal{P} : \nu \mbox{ minimizes } R(\nu |\rho) + \beta f_K(\nu)
\}. \nonumber 
\eea
Similarly, because of the large deviation upper bound for $L_n$ 
with respect to the microcanonical ensemble, it is natural
to define the set of microcanonical equilibrium macrostates
to be
\bea
\label{eqn:micromacrostates}
\mathcal{E}^{u,K} & \doteq & \{ \nu
\in \mathcal{P} : I^{u,K} (\nu) = 0 \} \\ 
& = & \{ \nu \in \mathcal{P}
: \nu \mbox{ minimizes } R(\nu | \rho) \ \mbox{subject to} \ f_K
(\nu) = u \}. \nonumber 
\eea
Each element $\nu$ in ${\cal E}_{\beta,K}$ and ${\cal E}^{u,K}$ has
the form $\nu = \nu_{-1}\delta_{-1} + \nu_0 \delta_0 + \nu_1 \delta_1$
and describes an equilibrium configuration of the model in the corresponding
ensemble.  For $j = -1, 0, 1$, $\nu_j$ gives the asymptotic relative
frequency of spins taking the value $j$.

In the next section we begin our study of the sets of equilibrium
macrostates for the BEG model by analyzing $\mathcal{E}_{\beta,K}$.


\section{Structure of the Set of Canonical Equilibrium Macrostates}
\beginsec

In this section, we give a complete description of the set
$\mathcal{E}_{\beta,K}$ of canonical equilibrium macrostates for all
values of $\beta$ and $K$.  In contrast to all other studies of the model, 
which fix $K$ and vary $\beta$, we analyze the structure of $\mathcal{E}_{\beta,K}$
by fixing $\beta$ and varying $K$.  
  As stated in Theorems \ref{thm:Ln2order}
and \ref{thm:Ln1order}, there exists a critical value of $\beta$,
denoted by $\beta_c$ and equal to $\log 4$, such that
$\mathcal{E}_{\beta,K}$ has two different forms for $\beta \leq
\beta_c$ and for $\beta > \beta_c$.  Specifically, for fixed $\beta \leq \beta_c$
$\, \ebk$ exhibits a continuous bifurcation as $K$ passes through a critical value
$K_c^{(2)}(\beta)$,
while for fixed $\beta > \beta_c$ $\, \ebk$ exhibits a discontinuous
bifurcation as $K$ passes through a critical value $K_c^{(1)}(\beta)$.
In Section 5 we show how to extrapolate
this information to information concerning the phase transition behavior of the canonical ensemble
for varying $\beta$:
a continuous, second-order phase transition for all fixed,
sufficiently large values of $K$ and  
a discontinuous, first-order phase transition for all 
fixed, sufficiently small values of $K$.

In terms of the uniform measure $\rho \doteq \frac{1}{3}(\delta_{-1} +
\delta_0 + \delta_1)$, we define 
\be
\label{eqn:absorbedprior}
\rho_\beta (d\omega_j) \doteq \frac{1}{Z(\beta)} \cdot \exp(-\beta
\omega_j^2) \, \rho(d\omega_j), 
\ee 
where $Z(\beta) \doteq \int_\Lambda
\exp(-\beta \omega_j^2) \, \rho(d\omega_j)$.
The next two theorems give the form of $\mathcal{E}_{\beta,K}$ for
$\beta \leq
\beta_c$ and for $\beta > \beta_c$.

\begin{thm}
\label{thm:Ln2order} \per
Define $\beta_c \doteq \log 4$ and fix $\beta \leq \beta_c$.  Let
$\rho_\beta$ be the measure defined in {\em(\ref{eqn:absorbedprior})}.
The following results hold.

{\em(a)} There exists a critical value $K_c^{(2)}(\beta) > 0$ such that 

\hsp {\em (i)} for $K \leq K_c^{(2)}(\beta)$, $\mathcal{E}_{\beta,K} = \{
\rho_\beta \}${\em ;}

\hsp {\em (ii)} for $K > K_c^{(2)}(\beta)$, there exist probability
measures $\nu^+ = \nu^+(\beta,K)$ and $\nu^- = \nu^-(\beta,K)$ such that
$\nu^+ \neq \nu^- \neq \rho_\beta$ and $\mathcal{E}_{\beta,K} = \{
\nu^+, \nu^- \}$.

{\em (b)} If we write $\nu^+ \in \mathcal{P}$ as
$\nu^+ = \nu_{-1}^+ \delta_{-1} + \nu_0^+
\delta_0 + \nu_1^+ \delta_1$, then $\nu^- =
\nu_1^+ \delta_{-1} + \nu_0^+ \delta_0 +
\nu_{-1}^+ \delta_1$.

{\em (c)} $\nu^+$ and $\nu^-$ are continuous functions of $K >
K_c^{(2)}(\beta)$, and both $\nu^+ \goto
\rho_\beta$ and $\nu^- \goto
\rho_\beta$ as $K \goto (K_c^{(2)}(\beta))^+$.
\end{thm}

Property (c) describes a continuous bifurcation in ${\cal E}_{\beta,K}$
as $K \goto (K_c^{(2)}(\beta))^+$.  This analogue of a second-order
phase transition explains the superscript 2 on
the critical value $K_c^{(2)}(\beta)$. 
The following theorem shows that for $\beta > \beta_c$ 
the set ${\cal E}_{\beta,K}$ undergoes a discontinuous bifurcation as
$K \goto (K_c^{(1)}(\beta))^+$.  This analogue of a first-order
phase transtion explains the superscript 1 on
the corresponding critical value $K_c^{(1)}(\beta)$.

\begin{thm}
\label{thm:Ln1order} \per
Define $\beta_c \doteq \log 4$ and fix $\beta > \beta_c$.  Let
$\rho_\beta$ be the measure defined in {\em(\ref{eqn:absorbedprior})}.
The following results hold.

{\em(a)} There exists a critical value $K_c^{(1)}(\beta) > 0$ such that

\hsp {\em (i)} for $K < K_c^{(1)}(\beta)$, $\mathcal{E}_{\beta,K} = \{
\rho_\beta \}${\em ;}

\hsp {\em (ii)} for $K = K_c^{(1)}(\beta)$, there exist probability
measures $\nu^+ = \nu^+(\beta,K_c^{(1)}(\beta))$ and $\nu^- =
\nu^-(\beta,K_c^{(1)}(\beta))$ such that $\nu^+ \neq \nu^- \neq
\rho_\beta$ and $\mathcal{E}_{\beta,K} = \{
\rho_\beta, \nu^+, \nu^- \}$.

\hsp {\em (iii)} For $K > K_c^{(1)}(\beta)$, there exist probability
measures $\nu^+ = \nu^+(\beta,K)$ and $\nu^- = \nu^-(\beta,K)$ such that
$\nu^+ \neq \nu^- \neq \rho_\beta$ and $\mathcal{E}_{\beta,K} = \{\nu^+,
\nu^- \}$.

{\em (b)} If we write $\nu^+ \in \mathcal{P}$ as
$\nu^+ = \nu_{-1}^+ \delta_{-1} + \nu_0^+
\delta_0 + \nu_1^+ \delta_1$, then $\nu^- =
\nu_1^+ \delta_{-1} + \nu_0^+ \delta_0 +
\nu_{-1}^+ \delta_1$.  

Since both $\nu^+(\beta,K) \neq \rho_\beta$ and $\nu^-(\beta,K) \neq
\rho_\beta$ for all $K \geq K_c^{(1)}(\beta)$, the bifurcation in 
${\cal E}_{\beta,K}$
at $K = K_c^{(1)}(\beta)$ is discontinuous.

\end{thm}

We prove Theorems \ref{thm:Ln2order} and \ref{thm:Ln1order} in
several steps.  In the first step, carried out in Section 3.1,
we absorb the noninteracting component
of the energy per particle into the product measure of the canonical
ensemble.
This reduces the model to a Curie-Weiss-type model, which can be
analyzed in terms of the empirical means $\sum_{j=1}^n \omega_j / n$.
The structure of the set of canonical equilibrium macrostates for this
Curie-Weiss-type model is analyzed in Section 3.2 for $\beta \leq \beta_c$
and in Section 3.3 for $\beta > \beta_c$.
Finally, in Section 3.4 we lift our results from the
level of the empirical means up to the level of the empirical measures using the
contraction principle, a main tool in the theory of large deviations.


\subsection{Reduction to the Curie-Weiss Model}

The first step in the proofs of Theorems \ref{thm:Ln2order} and
\ref{thm:Ln1order} is to rewrite the canonical ensemble
$P_{n,\beta,K}$ in the form of a Curie-Weiss-type model.  We do this by
absorbing the noninteracting component of the energy per particle
$h_{n,K}$
into the product measure of $P_{n,\beta,K}$.  Defining 
$S_n(\omega) = \sum_{j=1}^n \omega_j$, we write
\beas
\label{eqn:absorb}
P_{n,\beta,K}(d\omega) & \doteq & \frac{1}{Z_n(\beta,K)} \cdot \exp\!\left[ -n
\beta h_{n,K}(\omega) \right] P_n(d\omega) \\
& = & \frac{1}{Z_n(\beta,K)} \cdot \exp \!\left[ -n \beta
\left(\frac{\sum_{j=1}^n \omega_j^2}{n} - K \left(\frac{\sum_{j=1}^n \omega_j}{n} \right)^2
\right) \right] P_n(d\omega) \nonumber \\ 
& = & \frac{1}{Z_n(\beta,K)} \cdot \exp \!\left[ n \beta K
\left(\frac{S_n(\omega)}{n} \right)^2
\right] \prod_{j=1}^n \exp (-\beta \omega_j^2) \rho(d\omega_j) \nonumber
\\ 
& = & \frac{(Z(\beta))^n}{Z_n(\beta,K)} \cdot \exp\!\left[ n \beta K
\left(\frac{S_n(\omega)}{n} \right)^2
\right] P_n^\beta(d\omega). \nonumber
\eeas 
In this formula $Z(\beta) \doteq \int_\Lambda
\exp(-\beta \omega_j^2) \rho(d\omega_j)$ and $P_n^\beta$ is the
product measure on $\Lambda^n$ with one-dimensional marginals
$\rho_\beta$ defined in
(\ref{eqn:absorbedprior}).

We define 
\[
\tilde{Z}_n(\beta,K) \doteq \int_{\Lambda^n} \exp\!\left[ n \beta
\left(\frac{S_n}{n}\right)^2 \right] dP_n^\beta.
\]
Since $P_{N,\beta,K}$ is a probability measure, it follows that 
\[
\tilde{Z}_n(\beta,K) = \frac{Z_n(\beta,K)}{Z(\beta))^n}
\]
and thus that
\be
\label{eqn:rewritecanon}
P_{n,\beta,K}(d\omega) = \frac{1}{\tilde{Z}_n(\beta,K)} \cdot \exp\!\left[ n \beta K
\left(\frac{S_n(\omega)}{n} \right)^2 \right] P_n^\beta(d\omega). 
\ee

By expressing the
canonical ensemble in terms of the empirical means $S_n/n$, we
have reduced the BEG model to a Curie-Weiss-type
model.  Cram\'{e}r's Theorem \cite[Thm II.4.1]{Ellis} 
states that with respect to the product measure $P_n^\beta$,
$S_n/n$ satisfies the LDP on $\R$ with rate function 
\be
\label{eqn:cramerrate}
J_\beta (z) \doteq \sup_{t \in \R} \{tz - c_\beta (t) \}. 
\ee
In this formula 
$c_\beta$ is the cumulant generating function defined by
\bea
\label{eqn:cbeta}
c_\beta
(t) & \doteq & \log \int_\Lambda \exp (t\omega_1) \, \rho_\beta
(d\omega_1) \\ 
& = & \log \left[
\frac{1+e^{-\beta}(e^t+e^{-t})}{1+2e^{-\beta}} \right]. \nonumber 
\eea
$J_\beta$ is finite on the closed interval $[-1,1]$ and 
is differentiable on the open interval $(-1,1)$.
This function is expressed in (\ref{eqn:cramerrate}) as the
Legendre-Fenchel transform of the finite, convex, differentiable
function $c_\beta$.  
By the theory of these
transforms \cite[Thm.\ 25.1]{Rock}, \cite[Thm.\ VI.5.3(d)]{Ellis}, for each $z \in (-1,1)$
\be
\label{eqn:inverse}
J_\beta '(z) = (c_\beta ')^{-1}(z).
\ee 

From the LDP for $S_n/n$ with respect to $P_n^\beta$,
Theorem 2.4 in \cite{EHT1} gives the LDP for $S_n/n$ with respect to 
the canonical ensemble written in the form (\ref{eqn:rewritecanon}).

\begin{thm}\per
\label{thm:ldpsn}
With respect to the canonical ensemble $P_{n,\beta,K}$ written 
in the form {\em(\ref{eqn:rewritecanon})}, the empirical means
$S_n/n$ satisfy the LDP on $[-1,1]$ with rate function
\be
\label{eqn:empmeanrate}
\tilde{I}_{\beta,K} \doteq J_\beta (z) - \beta K z^2 - \inf_{t \in \R}
\{J_\beta (t) - \beta K t^2 \}.
\ee
\end{thm}

In Section 2 the canonical ensemble for the BEG model was expressed in terms of
the empirical measures $L_n$.  The corresponding
set $\E_{\beta,K}$ of canonical equilibrium macrostates 
 was defined as the set of
probability measures $\nu \in \mathcal{P}$ for which the rate function
$I_{\beta,K}$ in the associated LDP 
satisfies $I_{\beta,K} (\nu) = 0$ [see (\ref{eqn:canonmacrostates})]. 
By contrast, in (\ref{eqn:rewritecanon}) the canonical ensemble is expressed
in terms of the empirical means $S_n/n$.  We now consider the set
$\tilde{\mathcal{E}}_{\beta,K}$ of canonical equilibrium macrostates for the BEG
model expressed in terms of the empirical means.
The last theorem makes it natural to define 
$\tilde{\mathcal{E}}_{\beta,K}$ as the set of $z \in [-1,1]$
for which the rate function in that theorem 
satisfies $\tilde{I}_{\beta,K} (z) = 0$.  Since $z$ is a zero of this
rate function if and only if $z$ minimizes $J_\beta (z) - \beta K z^2$,
we have 
\be
\label{eqn:Snmacrostate}
\tilde{\mathcal{E}}_{\beta,K} \doteq  \{z \in [-1,1] : z
\mbox{ minimizes } \ J_\beta (z) - \beta K z^2 \}.
\ee
As we will see in Theorem \ref{thm:one-to-one}, each $z \in \tilde{{\cal E}}_{\beta,K}$
equals the mean of a corresponding measure in $\nu \in {\cal E}_{\beta,K}$.
Thus, each $z \in \tilde{{\cal E}}_{\beta,K}$ describes an equilibrium configuration
of the model in terms of the specific magnetization, or the asymptotic average spin per site.

Although $J_\beta(z)$ can be computed explicitly, the expression is
messy. Instead, we use an alternative characterization of $\tebk$ given in the
next proposition to determine the points in that set.
This proposition is a special case of a general result to
be presented in \cite{CET}.

\begin{prop}\per
\label{lemma:equivmin}
For $z \in \R$ define 
\be
\label{eqn:GbetaK} G_{\beta,K}(z) \doteq \beta K z^2 - 
c_\beta(2\beta K z).
\ee  
Then for each $\beta > 0$ and $K > 0$ 
\be
\label{eqn:altmin}
\min_{|z| \leq 1} \{J_\beta (z) - \beta K z^2 \} =  \min_{z \in \R}
\{ G_{\beta,K}(z)\}. 
\ee
In addition, the global minimum points of $J_\beta (z) - \beta K z^2$
coincide with the global minimum points of 
$G_{\beta,K}$.  As a consequence,
\be
\label{eqn:Snmacrostate2}
\tilde{\mathcal{E}}_{\beta,K} = \{ z \in \R : z  \mbox{ minimizes }
G_{\beta,K}(z) \}.  
\ee
\end{prop}

\noi {\bf Proof.} The finite, convex function $f(z) \doteq c_\beta(2\beta K z)/2\beta K$
has the Legendre-Fenchel transform
\[
f^*(z) = \sup_{x \in \R} \{xz - f(x)\} = 
\left\{ \begin{array}{ll} J_\beta(z)/2\beta K & \mbox{ for } |z| \leq 1 \\
\infty & \mbox{ for } |z| > 1.
\end{array}
\right.  
\]
We prove the proposition by showing the following three steps.
\begin{enumerate}
\item $\sup_{z \in \R} \{f(z) - z^2/2\} =
\sup_{|z| \leq 1}\{z^2/2 - f^*(z)\}$.
\item Both suprema in Step 1 are attained, the first for some $z \in \R$
and the second for some $z \in (-1,1)$.
\item The global maximum points of $f(z) - z^2/2$ coincide with 
the global maximum points of $z^2/2 - f^*(z)$.
\end{enumerate}
The proof uses three properties of Legendre-Fenchel transforms.
\begin{enumerate}
\item For all $z \in \R$, $f^{**}(z) = (f^*)^*(z)$ equals 
$f(z)$ \cite[Thm.\ VI.5.3(e)]{Ellis}.
\item If for some $x \in \R$ and $z \in \R$ we have $z = f'(x)$, then $f(x) + f^*(z) = xz$
\cite[Thm.\ 25.1]{Rock}, \cite[Thm.\ VI.5.3(c)]{Ellis}.
In particular, if $z = x$, then $f(x) + f^*(x) = x^2$.  
\item If there exists $x \in (-1,1)$ and $y \in \R$ such that 
\be
\label{eqn:fstarz}
f^*(z) \geq f^*(x) + y(z-x) \mbox{ for all } z \in [-1,1],
\ee
then $y = (f^*)'(x)$ \cite[Thm.\ 25.1]{Rock}. 
Hence by properties 1 and 2 
\[
f^*(x) + f^{**}(y) = f^*(x) + f(y) = xy.
\]
In particular, if (\ref{eqn:fstarz}) is valid with $y = x$, then $f(x) + f^*(x) = x^2$.  
\end{enumerate}

Step 1 in the proof is a special case of Theorem C.1 in \cite{EE}.  For completeness,
we present the straightforward proof.  Let $M = \sup_{z \in \R}\{f(z) - z^2/2\}$.  
Since for any $|z| \leq 1$ and $x \in \R$
\[
f^*(z) + M \geq xz - f(x) + M \geq xz - x^2/2,
\]
we have
\[
f^*(z) + M \geq \sup_{x \in \R} \{xz - x^2/2\} = z^2/2.
\]
It follows that $M \geq z^2/2 - f^*(z)$ and thus that
$M \geq \sup_{|z| \leq 1}\{z^2/2 - f^*(z)\}$.
To prove the reverse inequality, let $N = \sup_{|z| \leq 1}\{z^2/2 - f^*(z)\}$.
Then for any $z \in \R$ and $|x| \leq 1$
\[
z^2/2 + N \geq xz - x^2/2 + N \geq xz - f^*(x).
\]
Since $f^*(x) = \infty$ for $|x| > 1$, it follows from property 1 that
\[
z^2/2 + N \geq \sup_{|x| \leq 1}\{xz - f^*(x)\} = f(z)
\]
and thus that $N \geq \sup_{z \in \R}\{f(z) - z^2/2\}$.
This completes the proof of step 1.

Since $f(z) \sim |z|$ as $z \goto \infty$, $f(z) - z^2/2$ attains its
supremum over $\R$.  Since $z^2/2 - f^*(z)$ is continuous and 
$\lim_{|z| \goto 1} (f^*)'(z) = \infty$, $z^2/2 - f^*(z)$ attains its
supremum over $[-1,1]$ in the open interval $(-1,1)$.  This completes
the proof of step 2.

We now prove that the global maximum points of the two functions coincide.
Let $x$ be any point in $\R$ at which $f(z) - z^2/2$ attains its supremum.
Then $x = f'(x)$, and so by the second assertion in property 2 $f(x) + f^*(x) = x^2$.  
The point 
$x$ lies in $(-1,1)$ because the range of $f'(z) = c'_\beta(2\beta K z)$
equals $(-1,1)$.  Step 1 now implies that 
\beas
\sup_{z \in \R}\{f(z) - z^2/2\} & = & f(x) - x^2/2 \\
& = & x^2/2 - f^*(x)= \sup_{|z| \leq 1}\{z^2/2 - f^*(z)\}. 
\eeas
We conclude that $z^2/2 - f^*(z)$ attains its supremum at $x \in (-1,1)$.

Conversely, let $x$ be any point in $(-1,1)$ at which $z^2/2 - f^*(z)$
attains its supremum.  Then for any $z \in [-1,1]$
\[
x^2/2 - f^*(x)  \geq z^2/2 - f^*(z).
\]
It follows that for any $z \in [-1,1]$
\[
f^*(z) \geq f^*(x) + (z^2 - x^2)/2 \geq f^*(x) + x(z-x).
\]
The second assertion in property 3 implies that $f^*(x) + f(x) = x^2$, and
in conjunction with step 1 this in turn implies that
\beas
\sup_{|z| \leq 1}\{z^2/2 - f^*(z)\} & = & x^2/2 - f^*(x) \\
& = & f(x) - x^2/2 = \sup_{z \in \R}\{f(z) - z^2/2\}.
\eeas
We conclude that $f(z) - z^2/2$ attains its supremum at $x$.  This
completes the proof of the proposition.
 \ \ink

\vsp Proposition \ref{lemma:equivmin} states that
$\tilde{\mathcal{E}}_{\beta,K}$ consists of the global minimum points
of $G_{\beta,K}(z) \doteq \beta K z^2 - c_\beta(2 \beta K z)$.  In order to simplify the minimization problem, we
make the change of variables $z \goto z/2\beta K$ in $G_{\beta,K}$,
obtaining the new function
\be
\label{eqn:FbetaK}
F_{\beta,K}(z) \doteq G_{\beta,K}(z/2\beta K) = \frac{z^2}{4\beta K} - c_\beta (z).  
\ee
Proposition \ref{lemma:equivmin} gives the alternative
characterization of $\tilde{\mathcal{E}}_{\beta,K}$ to be 
\be
\label{eqn:EintermsofF}
\tilde{\mathcal{E}}_{\beta,K} = \left\{ \frac{w}{2\beta K} \in \R : w
\ \mbox{minimizes} \ F_{\beta,K} (w) \right\}.  
\ee
We use
$F_{\beta,K}$ to analyze 
$\tilde{\mathcal{E}}_{\beta,K}$ because the second term of $F_{\beta,K}$
contains only the parameter $\beta$ while both terms in $G_{\beta,K}$
contain both parameters $\beta$ and $K$.  In order to analyze the 
structure of $\tilde{{\cal E}}_{\beta,K}$, 
we take advantage of the
simpler form of $F_{\beta,K}$ by fixing $\beta$ and varying
$K$.  This innovation makes the analysis of
$\tilde{\mathcal{E}}_{\beta,K}$ much more efficient than in previous
studies.  Our goal is prove that the elements of
$\tilde{\mathcal{E}}_{\beta,K}$ change continuously with $K$ for all
sufficiently small values of $\beta$ [Thm.\ \ref{thm:Ln2order}]
and have a discontinuity at $K = K_c^{(1)}$ for all sufficiently large values of $\beta$
[Thm.\ \ref{thm:Ln1order}].

In order to determine the minimum points of $F_{\beta,K}$ and thus the
points in $\tilde{\mathcal{E}}_{\beta,K}$, we study its derivative
\be
\label{eqn:Fderiv}
F_{\beta,K}'(w) = \frac{w}{2 \beta K} - c_\beta'(w).
\ee
$F_{\beta,K}'(w)$ consists of
 a linear part $w/2\beta K$ and a nonlinear part $c_\beta'(w)$.  
According to Theorems \ref{thm:Ln2order} and \ref{thm:Ln1order}, 
${\cal E}_{\beta,K}$ exhibits a continuous bifurcation in $K$
when $\beta \leq \beta_c$ and a discontinuous bifurcation in $K$
where $\beta_c \doteq \log 4$.  
As we will see in Sections 3.2
and 3.3, the basic mechanism
underlying this change in the bifurcation behavior of 
${\cal E}_{\beta,K}$ is the change in the 
concavity behavior of $c_\beta'(w)$ 
for $\beta \leq \beta_c$ versus $\beta >
\beta_c$, which is the subject of the next theorem.  
A related phenomenon was observed in \cite[Thm.\ 1.2(b)]{EMN}
and in \cite[Thm.\ 4]{ENO} in the context of work on the 
Griffiths-Hurst-Sherman correlation inequality for models
of ferromagnets; this inequality is used to show the concavity 
of the specific magnetization as a function of the external field.

\begin{thm} \per
\label{thm:concavity}
Define $\beta_c \doteq \log 4$ and for $\beta > \beta_c$ define 
\be
\label{eqn:wcrit}
w_c(\beta) \doteq \cosh^{-1} \! \left( \frac{1}{2} e^\beta -
4e^{-\beta}  \right) \geq 0.  
\ee 
Then the following hold.

(a) For $\beta \leq \beta_c$, $c_\beta'(w)$ is strictly concave  for
$w > 0$.

(b) For $\beta > \beta_c$,
$c_\beta'(w)$ is strictly convex for $0 < w < w_c(\beta)$ \
and $c_\beta'(w)$ is strictly concave for $w > w_c(\beta)$.
\end{thm}

\noi {\bf Proof.} (a) We show
that for all $\beta \leq \beta_c$, $c_\beta '''(w) < 0$ for all $w >
0$.  A short calculation yields 
\bea
\label{eqn:ctripleprime}
c_\beta '''(w) & = & \frac{[e^{-\beta}(e^w - e^{-w})][1-e^{-\beta}(e^w
+ e^{-w})-8e^{-2\beta}]}{[1+e^{-\beta}(e^w + e^{-w})]^3}  \\
& = & \frac{[2e^{-\beta}\sinh w][1-2e^{-\beta}\cosh w -
8e^{-2\beta}]}{[1 + 2e^{-\beta}\cosh w]^3}. \nonumber 
\eea 
Since
$2e^{-\beta}\sinh w$ and $1 + 2e^{-\beta}\cosh w$ are 
positive for $w > 0$, $c_\beta '''(w) < 0$ for $w > 0$ if and only if
\[
1-2e^{-\beta}\cosh w -8e^{-2\beta} < 0 \ \mbox{ for } w > 0.
\]
The inequality $\cosh w > 1$ for $w > 0$ implies that 
\[ 
[1-2e^{-\beta}\cosh w -8e^{-2\beta}] < [1-2e^{-\beta} -8e^{-2\beta}] =
(1 - 4e^{-\beta})(1+2e^{-\beta}) \mbox{ for all } w > 0.
\]
Therefore, for all $\beta \leq \log 4$, $c_\beta '''(w) < 0$ for $w >
0$.

(b) Fixing $\beta > \beta_c$, we determine the critical value
$w_c(\beta)$ such that $c_\beta'(w)$ is strictly convex for $0 < w <
w_c(\beta)$ and strictly concave for $w > w_c(\beta)$.  From the
expression for $c_\beta '''(w)$ in (\ref{eqn:ctripleprime}), $c_\beta
'''(w) > 0$ for $w > 0$ if and only if $(1-2e^{-\beta}\cosh w
-8e^{-2\beta}) > 0$ for $w > 0$.
Therefore $c_\beta'(w)$ is strictly convex for
\[
0 < w < \cosh^{-1} \! \left( \frac{1}{2} e^\beta - 4e^{-\beta} \right).
\]
On the other hand, since $c_\beta '''(w) < 0$ for $w > 0$ if and only if
$(1-2e^{-\beta}\cosh w -8e^{-2\beta}) < 0$ for $w > 0$, we conclude that
$c_\beta'(w)$ is
strictly concave for
\[
w > \cosh^{-1} \left( \frac{1}{2} e^\beta - 4e^{-\beta} \right).
\]
This completes the proof of part (b). \ \ink

\skp
The concavity description of $c_\beta '$ stated in Theorem
\ref{thm:concavity} allows us to find the global minimum points of
$F_{\beta,K}$ and thus the points in $\tilde{\mathcal{E}}_{\beta,K}$
for all values of the parameters $\beta$ and $K$.  We carry this out
in the next two sections, first for $\beta \leq \beta_c$ and then for
$\beta > \beta_c$.  In Section 3.4 we use this information to give
the structure of the set ${\cal E}_{\beta,K}$ of canonical equilibrium
macrostates defined in (\ref{eqn:canonmacrostates}).


\subsection{Description of
\boldmath $\tilde{\mathcal{E}}_{\beta,K}$ \unboldmath
for \boldmath${\beta \leq \beta_c}$\unboldmath}

In Theorem \ref{thm:Ln2order} we stated the structure of the set
$\mathcal{E}_{\beta,K}$ of canonical equilibrium macrostates for the BEG
model with respect to the empirical measures when $\beta \leq \beta_c \doteq \log 4$.
The main theorem in this section, Theorem \ref{thm:secondorder}, does
the same for the set 
$\tilde{\mathcal{E}}_{\beta,K}$, which has been shown to have the
alternative characterization
\be
\label{eqn:EintermsofF2}
\tilde{\mathcal{E}}_{\beta,K} = \left\{ \frac{w}{2\beta K} \in \R : w
\ \mbox{minimizes} \ F_{\beta,K} (w) \right\}. 
\ee
We recall that $F_{\beta,K}(w) \doteq w^2/4\beta K - c_\beta(w)$, where
$c_\beta$ is defined in (\ref{eqn:cbeta}).  In Section 3.4 we use the 
fact that
there exists a one-to-one correspondence between
$\mathcal{E}_{\beta,K}$ and $\tilde{\mathcal{E}}_{\beta,K}$
to fully describe the latter set for all $\beta \leq \beta_c$ 
and $K > 0$.

According to Theorem \ref{thm:concavity}, for
$\beta \leq \beta_c$, $c_\beta '$ is strictly concave for
$w > 0$.  As a result, the study of $\tilde{\mathcal{E}}_{\beta,K}$ is
similar to the study of the equilibrium macrostates for the classical
Curie-Weiss model as given in Section IV.4 of \cite{Ellis}.  Following
the discussion
in that section, we use a graphical
argument to motivate the continuous bifurcation 
exhibited by
$\tilde{\mathcal{E}}_{\beta,K}$ for $\beta \leq \beta_c$.  A detailed proof
is given in Section 2.3.2 of \cite{O}.

Minimum points of $F_{\beta,K}$ satisfy $F_{\beta,K}'(w) = 0$, which can
be rewritten as 
\be
\label{eqn:Fderiv2}
\frac{w}{2\beta K} = c_\beta '(w).  
\ee 
Since the slope of the
function $w \mapsto w/2\beta K$ is $1/2\beta K$, the nature of the
solutions of (\ref{eqn:Fderiv2}) depends on whether
\[
c_\beta ''(0) \leq \frac{1}{2\beta
K} \hsp \mbox{or} \hsp 0 < \frac{1}{2\beta K} < c_\beta ''(0).
\]

Define the critical point
\be
\label{eqn:2ndordercrit}
K_c^{(2)}(\beta) \doteq \frac{1}{2\beta c_\beta ''(0)} = \frac{1}{4
\beta e^{-\beta}} + \frac{1}{2\beta}.  
\ee
We use the same notation here as for the critical value in Theorem
\ref{thm:Ln2order} because, as we will later prove, the continuous
bifurcation in $K$ exhibited by both sets
$\mathcal{E}_{\beta,K}$ and $\tilde{\mathcal{E}}_{\beta,K}$ occur at the
same point $K_c^{(2)}(\beta)$ defined in (\ref{eqn:2ndordercrit}).

We illustrate the minimum points of $F_{\beta, K}$ graphically in Figure
1 for $\beta = 1$.  For three ranges of
values of $K$ this figure depicts
 the two components of $F_{\beta,K}'$: the linear component $w/2\beta
K$ and the nonlinear component $c_\beta '(w)$.  Figure 1(a) corresponds to $0 < K < K_c^{(2)}(\beta)$.
Since from (\ref{eqn:2ndordercrit}) $c_\beta ''(0) = 1/2 \beta
K_c^{(2)}(\beta)$, for $0 < K < K_c^{(2)}(\beta)$ the two components of
$F_{\beta,K}'$ intersect at only the origin, and thus $F_{\beta,K}$ has a
unique global minimum point at $w=0$.  Figure 1(b) corresponds to 
$K = K_c^{(2)}(\beta)$.  In this case the two components of $F_{\beta,K}'$
are tangent at the origin, and again $F_{\beta,K}$ has a unique
global minimum point at $w = 0$.  Figure 1(c) corresponds to $K >
K_c^{(2)}(\beta)$.  For such $K$ the global minimum points of $F_{\beta,K}$
are symmetric nonzero points $w = \pm \tilde{w} (\beta,K)$,
$\tilde{w}(\beta,K) > 0$.  In addition, for $K >
K_c^{(2)}(\beta)$, $\tilde{w}(\beta,K)$ is a continuous function and
as $K \goto (K_c^{(2)}(\beta))^+$, $ \tilde{w} (\beta,K)$ converges to
$0$.  As a result, we conclude that $\tilde{\mathcal{E}}_{\beta,K}$
exhibits a continuous bifurcation with respect to $K$.

\begin{figure}[ht]
\begin{center}
\epsfig{file=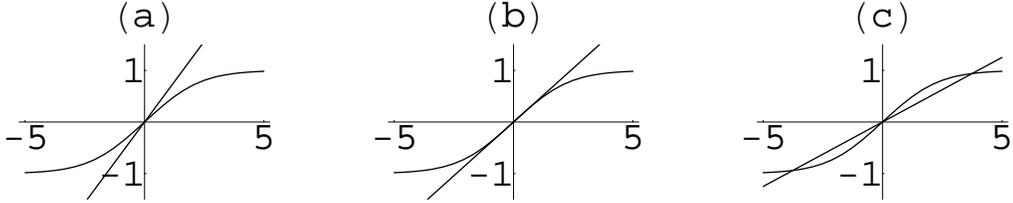,width=15cm}
\caption{ \small Continuous bifurcation for $\beta = 1$. (a) $K
< K_c^{(2)}(\beta)$, (b) $K = K_c^{(2)}(\beta)$, (c) $K > K_c^{(2)}(\beta)$ }
\end{center}
\end{figure} 

Figures 1(a) and 1(c) give similar information as Figures IV.3(b) and IV.3(d)
in \cite{Ellis}, 
which depict the phase transition in the Curie-Weiss model.  In these two sets
of figures the functions being graphed are Legendre-Fenchel transforms of each other.

Figures 1(a)--(c) motivate the following theorem 
concerning the continuous bifurcation with respect to $K$ exhibited by
$\tilde{\mathcal{E}}_{\beta,K}$ in the BEG model.  It is proved 
in Theorem 2.3.6 of \cite{O}.  The positive quantity $\tz(\beta,K)$ in parts 
(b) and (c) of the next theorem equals $\tw(\beta,K)/2\beta K$, where
$\tw(\beta,K)$ is the positive global minimum point of $F_{\beta,K}$
for $K > K_c^{(1)}(\beta)$.

\begin{thm}
\label{thm:secondorder} \per
Define $\beta_c \doteq \log 4$ and fix $\beta \leq \beta_c$.  Define
$\tilde{\mathcal{E}}_{\beta,K}$ by {\em(\ref{eqn:Snmacrostate})} and the
critical value
$K_c^{(2)}(\beta)$ by {\em(\ref{eqn:2ndordercrit})}.  The following
results hold.

{\em(a)} For $K \leq K_c^{(2)}(\beta)$,
$\tilde{\mathcal{E}}_{\beta,K} = \{0\}$.

{\em(b)} For $K > K_c^{(2)}(\beta)$, there exists a positive number
$\tilde{z}(\beta,K)$ such that
$\tilde{\mathcal{E}}_{\beta,K} = \{\pm \tilde{z} (\beta,K) \}$.

{\em(c)} $\tilde{z} (\beta,K)$ is
a strictly increasing continuous function of $K > K_c^{(2)}(\beta)$, and
$\tilde{z} (\beta,K) \goto 0$ as $K \goto (K_c^{(2)}(\beta))^+$. 
\end{thm}

This theorem completes our description of the continuous
bifurcation exhibited by $\tilde{\mathcal{E}}_{\beta,K}$
for $\beta \leq \beta_c$.  In the next section we describe
the discontinuous bifurcation exhibited by 
$\tilde{\mathcal{E}}_{\beta,K}$
for $\beta$ in the complementary region $\beta > \beta_c$.


\subsection{Description of
\boldmath $\tilde{\mathcal{E}}_{\beta,K}$ \unboldmath for
\boldmath${\beta > \beta_c}$\unboldmath}

In Theorem \ref{thm:Ln1order} we gave the structure of the set
$\mathcal{E}_{\beta,K}$ of canonical equilibrium macrostates for the BEG
model with respect to the empirical measures when $\beta > \beta_c$.
The main theorem in this section, Theorem \ref{thm:firstorder}, does the
same for the set 
$\tilde{\mathcal{E}}_{\beta,K}$, which has been shown to have the
alternative characterization
\be
\label{eqn:EintermsofF3}
\tilde{\mathcal{E}}_{\beta,K} = \left\{ \frac{w}{2\beta K} \in \R : w
\mbox{ minimizes } F_{\beta,K} (w) \right\}. 
\ee 
As in Section 3.2, $F_{\beta,K}(w) \doteq w^2/4\beta K - c_\beta(w)$,
where $c_\beta$ is defined in (\ref{eqn:cbeta}).  In the next section we
use the fact that there exists a one-to-one correspondence between 
$\mathcal{E}_{\beta,K}$ and $\tilde{\mathcal{E}}_{\beta,K}$ to fully
describe the latter set for all $\beta > \beta_c$ and $K > 0$.

Minimum points of $F_{\beta,K}$ satisfy the equation
\be
\label{eqn:Fderiv3}
F_{\beta,K}'(w) = \frac{w}{2\beta K} - c_\beta '(w) = 0.  
\ee 
In contrast to the previous section, where for $\beta \leq \beta_c$
$\, c_\beta '$ is strictly concave for $w > 0$, part (b) of  Theorem
\ref{thm:concavity} states that for $\beta > \beta_c$ there exists $w_c
= w_c(\beta) > 0$ such that $c_\beta '$ is strictly convex for $w
\in (0, w_c)$ and strictly concave for $w > w_c$.  As a result, for
$\beta > \beta_c$ we are no longer in the situation of the classical Curie-Weiss 
model for which the bifurcation with respect to $K$ is continuous.  Instead, for $\beta >
\beta_c$, as $K$ increases through the critical value $K_c^{(1)}
(\beta)$, $\tilde{\mathcal{E}}_{\beta,K}$ exhibits a discontinuous
bifurcation.

While the discontinuous bifurcation exhibited by
$\tilde{\mathcal{E}}_{\beta,K}$ for $\beta > \beta_c$ is easily observed
graphically, the full analytic proof is considerably
more complicated than in the case $\beta \leq \beta_c$.
As in the previous section, we will motivate this discontinuous
bifurcation via a graphical argument, referring the
reader to Section 2.3.3 of \cite{O} for details.

For $\beta > \beta_c$ we divide the range of the positive parameter $K$ into
three intervals separated by the values $K_1 = K_1(\beta)$ and $K_2
=K_2(\beta)$.  $K_1$ is defined to be the unique value of $K$ such that
the line $w/2\beta K$ is tangent to the curve $c_\beta '$ at some positive
point $w_1 = w_1(\beta)$.  The existence and uniqueness of $K_1$ and 
$w_1$ are proved in Lemma 2.3.8 in \cite{O}.  $K_2$
is defined to be the value of $K$ such that the slopes of the line
$w/2\beta K$ and the curve $c_\beta '$ at $w=0$ agree.  Specifically,  
\be
\label{eqn:K2}
K_2 \doteq \frac{1}{2\beta c_\beta ''(0)} = \frac{1}{4 \beta e^{-\beta}}
+ \frac{1}{2\beta}.  
\ee
Figure 2 represents graphically the values of $K_1$ and
$K_2$ for $\beta = 4$.  That figure exhibits $K_1 < K_2$; in Lemma
2.3.9 in \cite{O} it is proved that this inequality holds for all
$\beta > \beta_c$.

\begin{figure}[ht]
\begin{center}
\epsfig{file=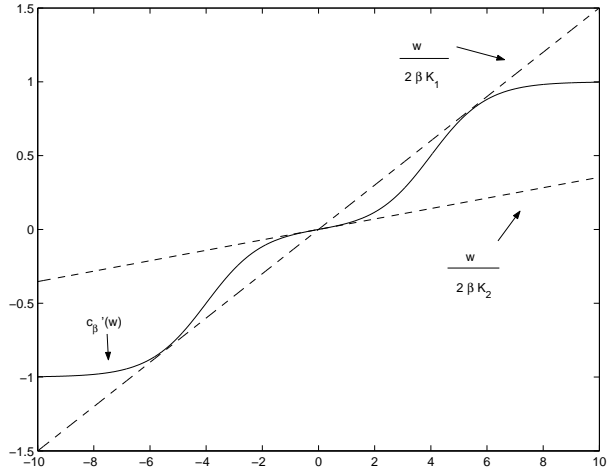,width=8cm}
\caption{ \small Graphical representation of the values $K_1$ and
$K_2$ for $\beta = 4$. }
\end{center}
\end{figure}

In each of Figures 3--7, for fixed $\beta > \beta_c$ and for different ranges of
values of $K > 0$, the first graph $(a)$ depicts the two components of
$F_{\beta,K}'$: the linear component $w/2\beta K$ and the nonlinear component
$c_\beta '$.  The second graph $(b)$ shows the corresponding graph of
$F_{\beta,K}$.  In these figures the following values of $\beta$ 
were used: $\beta = 4$ in Figures 3, 5, 6, 7 and $\beta = 2.8$ in Figure 4.

As we see in Figure 3, for $K \in (0, K_1]$ the linear
term intersects the nonlinear term at only the origin and thus
$F_{\beta,K}$ has a unique global minimum point at $w = 0$.  Since
$\beta$ is fixed, the graph of the nonlinear term $c_\beta'$ also
remains fixed.  As the value of $K$ increases, the slope of the
linear term $w/2\beta K$ decreases, leading to the discontinuous
bifurcation in ${\cal E}_{\beta,K}$ with respect to $K$.

The graph of $F_{\beta, K}$ is depicted in Figure 4 for $K \in [K_2,
\infty)$.  We see that $F_{\beta,K}$ has two global minimum points at $w
= \pm
\tilde{w}(\beta, K)$, where $\tilde{w}(\beta, K)$ is positive.
Therefore, by
(\ref{eqn:EintermsofF3}), for $0 < K \leq K_1$ we have
$\tilde{\mathcal{E}}_{\beta,K} = \{0\}$ and for $K \geq K_2$ we have
$\tilde{\mathcal{E}}_{\beta,K} = \{ \pm \tilde{z}(\beta, K)
\}$, where $\tilde{z}(\beta,K) \doteq \tilde{w}(\beta, K)/2\beta K$ is
positive.

\begin{figure}[ht]
\begin{center}
\epsfig{file=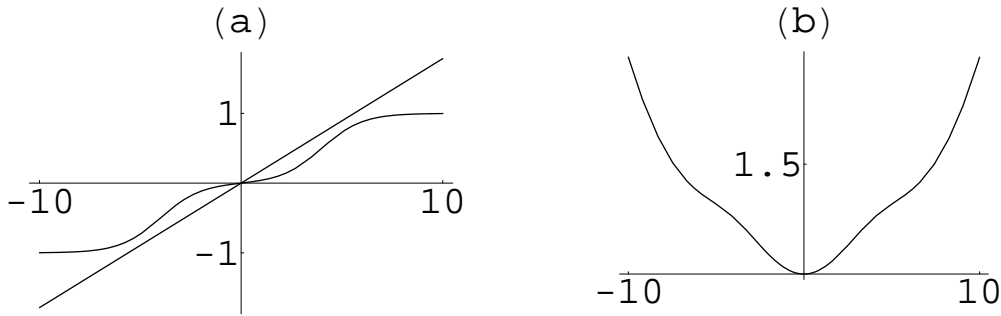,width=15cm}
\caption{ \small (a) Graph of two components of $F_{\beta,K}'$ and (b)
graph of $F_{\beta,K}$ for $K \in (0, K_1]$ }
\end{center}
\end{figure}

\begin{figure}[ht]
\begin{center}
\epsfig{file=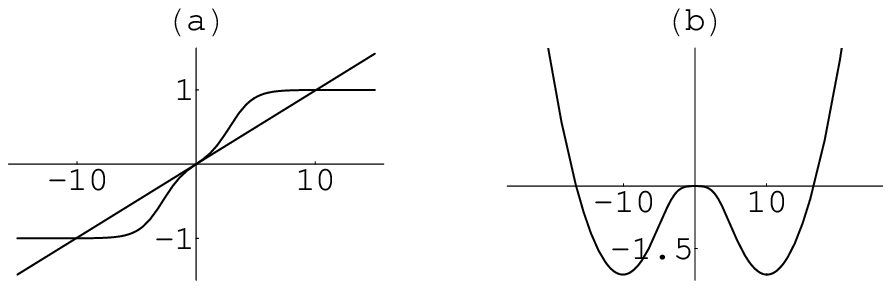,width=15cm}
\caption{ \small (a) Graph of two components of $F_{\beta,K}'$ and (b)
graph of $F_{\beta,K}$ for $K \geq K_2$ }
\end{center}
\end{figure}

Now suppose that $K \in (K_1, K_2)$.  In this region, there exists
$\tilde{w} (\beta,K) > 0$ such that $F_{\beta,K}$ has three local
minimum points at $w=0$ and $w = \pm \tilde{w} (\beta,K)$.  As we see in
Figure 5, for $K$ slighty greater than $K_1$, $F_{\beta,K}(0) <
F_{\beta,K}(\tilde{w} (\beta,K))$; as a result, the unique global
minimum point of $F_{\beta,K}$ is $w=0$.   On the other hand, we see in
Figure 6 that for $K$ slightly less than $K_2$, $F_{\beta,K}(0) >
F_{\beta,K}(\tilde{w} (\beta,K))$; as a result, the global minimum
points of $F_{\beta,K}$ are $w=\pm \tilde{w} (\beta,K)$.  
As $K$ increases over the interval $(K_1, K_2)$,
$F_{\beta,K}(0)$ increases and $F_{\beta,K}(\tilde{w} (\beta,K))$
decreases continuously and consequently, as Figure 7 reveals, there exists a critical value
$K_c^{(1)}(\beta)$ such that $F_{\beta,K_c^{(1)}(\beta)}(0) =
F_{\beta,K_c^{(1)}(\beta)}(\tilde{w} (\beta,K))$; as a result, the
global minimum points of $F_{\beta,K_c^{(1)}(\beta)}$ are $w=0$ and
$w=\pm \tilde{w} (\beta,K)$.

\begin{figure}[ht]
\begin{center}
\epsfig{file=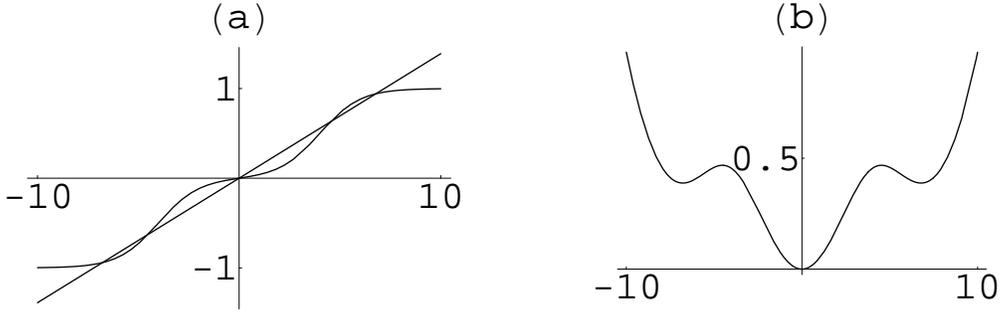,width=15cm}
\caption{ \small (a) Graph of two components of $F_{\beta,K}'$ and (b)
graph of $F_{\beta,K}$ for $K_1 < K < K_c^{(1)}(\beta)$ }
\end{center}
\end{figure}

\begin{figure}[ht]
\begin{center}
\epsfig{file=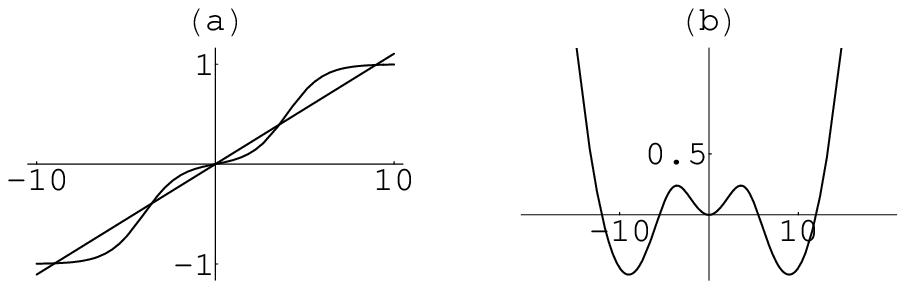,width=15cm}
\caption{ \small (a) Graph of two components of $F_{\beta,K}'$ and (b)
graph of $F_{\beta,K}$ for $K_c^{(1)}(\beta) < K < K_2$ }
\end{center}
\end{figure}

\begin{figure}[ht]
\begin{center}
\epsfig{file=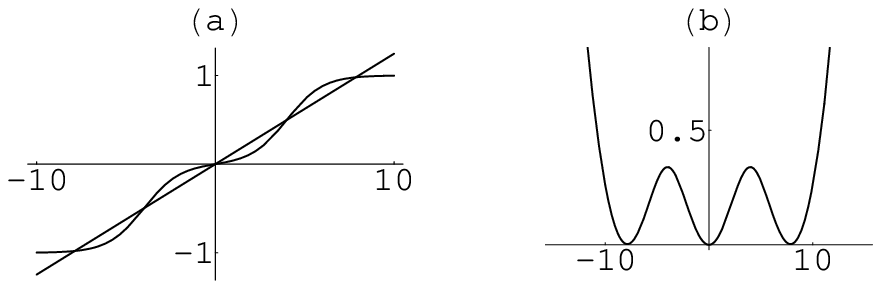,width=15cm}
\caption{ \small (a) Graph of two components of $F_{\beta,K}'$ and (b)
graph of $F_{\beta,K}$ for $K = K_c^{(1)}(\beta)$ }
\end{center}
\end{figure}

In conclusion, we have the following picture: for $K_1 < K <
K_c^{(1)}(\beta)$, $\tilde{\mathcal{E}}_{\beta,K} = \{0\}$; for $K =
K_c^{(1)}(\beta)$, $\tilde{\mathcal{E}}_{\beta,K} = \{0, \pm
\tilde{z}(\beta,K) \}$; and for $K > K_c^{(1)}(\beta)$,
$\tilde{\mathcal{E}}_{\beta,K} = \{\pm \tilde{z}(\beta,K) \}$.  Lastly,
since $\tilde{z}(\beta,K_c^{(1)}(\beta))$ is positive, the
bifurcation exhibited by $\tilde{\mathcal{E}}_{\beta,K}$ at $K =
K_c^{(1)}(\beta)$ is discontinuous.  The same notation
$K_c^{(1)}(\beta)$ is used here to denote the critical value for the
bifurcation exhibited by $\tilde{\mathcal{E}}_{\beta,K}$ as we used
to denote the critical value for the bifurcation exhibited by
$\mathcal{E}_{\beta,K}$ stated in Theorem \ref{thm:Ln1order}.  This is
appropriate because the two critical values are equal.

The discontinuous bifurcation exhibited by
$\tilde{\mathcal{E}}_{\beta,K}$ for $\beta > \beta_c$ 
is described in the following theorem.  This result corresponds
to Theorem 2.3.7 in \cite{O}, where a complete proof is presented.
The quantity $\tz(\beta,K)$ in parts (b)--(d) of the next theorem
equals $\tw(\beta,K)/2\beta K$, where $\tw(\beta,K)$ is the positive
global minimum point of $F_{\beta,K}$ for $K \geq K_c^{(1)}(\beta)$.

\begin{thm}
\label{thm:firstorder} \per
Define the set $\tilde{\mathcal{E}}_{\beta,K}$ by
{\em (\ref{eqn:Snmacrostate})}.  For a fixed $\beta > \beta_c \doteq \log
4$, there exists a critical value $K_c^{(1)}(\beta)$ such that the the
following
hold.

{\em(a)} For $K < K_c^{(1)}(\beta)$,
$\tilde{\mathcal{E}}_{\beta,K} = \{0\}$.

{\em(b)} For $K = K_c^{(1)}(\beta)$, $\tilde{\mathcal{E}}_{\beta,K} =
\{0,\pm \tilde{z}(\beta,K) \} \ \mbox{where} \ \tilde{z} (\beta,K) >
0$.

{\em(c)} For $K > K_c^{(1)}(\beta)$, $\tilde{\mathcal{E}}_{\beta,K} =
\{\pm
\tilde{z}(\beta,K) \} \ \mbox{where} \ \tilde{z} (\beta,K) > 0$.

{\em(d)} For $K \geq K_c^{(1)}(\beta)$, $\tilde{z} (\beta,K)$
is a strictly increasing continuous function and $\tilde{z}
(\beta,K_c^{(1)}(\beta)) > 0$.  Therefore,
$\tilde{\mathcal{E}}_{\beta,K}$ exhibits a discontinuous bifurcation.
\end{thm}

Together, Theorems \ref{thm:secondorder} and \ref{thm:firstorder} give a
full description of the set
$\tilde{\mathcal{E}}_{\beta,K}$ for all values of $\beta$ and $K$.  In
the next section, we use the contraction 
principle to lift our results concerning the structure of the set
$\tilde{\mathcal{E}}_{\beta,K}$ up to the level of the empirical
measures, making use of a
one-to-one correspondence between the points in the two sets
$\tilde{\mathcal{E}}_{\beta,K}$ and $\mathcal{E}_{\beta,K}$ of canonical
equilibrium macrostates.


\subsection{One-to-One Correspondence Between
\boldmath $\mathcal{E}_{\beta,K}$ \unboldmath and
\boldmath $\tilde{\mathcal{E}}_{\beta,K}$ \unboldmath}

We start by recalling the definitions of the sets
$\mathcal{E}_{\beta,K}$ and $\tilde{\mathcal{E}}_{\beta,K}$:
\be
\label{eqn:canonmacrostates2}
\mathcal{E}_{\beta,K} = \{ \nu \in \mathcal{P}: \nu \mbox{ minimizes } 
R(\nu | \rho) + \beta f_K(\nu) \}
\ee 
and 
\be
\label{eqn:Snmacrostate22}
\tilde{\mathcal{E}}_{\beta,K} = \{ z \in [-1,1] : z \mbox{ minimizes }
J_\beta(z) - \beta K z^2 \}.  
\ee 
In the definition of
$\mathcal{E}_{\beta,K}$, $R(\mu | \rho)$ is the relative entropy of
$\mu$ with respect to $\rho = \frac{1}{3}(\delta_{-1} + \delta_0 +
\delta_1)$ and $f_K (\mu)$ is the 
function defined in (\ref{eqn:interactfcn}).  In the definition of
$\tilde{\mathcal{E}}_{\beta,K}$, $J_\beta$ is the Cram\'{e}r rate
function defined in (\ref{eqn:cramerrate}).  We now state the
one-to-one correspondence between the
points in $\tilde{\mathcal{E}}_{\beta,K}$ and the points in
$\mathcal{E}_{\beta,K}$.  According to Theorems \ref{thm:secondorder}
and
\ref{thm:firstorder}, $\tilde{\mathcal{E}}_{\beta,K}$
consists of either $1,2$ or $3$ points.
\begin{thm}
\label{thm:one-to-one} \per
Fix $\beta > 0$ and $K > 0$ and suppose that
$\tilde{\mathcal{E}}_{\beta,K} = \{z_k\}_{k=1}^r$, $r=1,2 \ \mbox{or}
\ 3$. Define $\nu_k, k=1,...,r$, to be measures in $\mathcal{P}$ with
densities 
\be
\label{eqn:macromeasure}
\frac{d\nu_k}{d\rho_\beta} (y) \doteq \exp (t_k y) \cdot
\frac{1}{\displaystyle \int_\Lambda \exp (t_k y) \rho_\beta (dy)},
\ee
where $t_k$ is chosen such that $\int_\Lambda y \, \nu_k (dy) =
z_k$. Then for each $k = 1,...,r$, $t_k$ exists and is unique, and
$\mathcal{E}_{\beta,K}$ consists of the unique elements $\nu_k, k =1, \ldots, r$.  
Furthermore, $t_k = 2 \beta K z_k$ for $k = 1,...,r$.
\end{thm}

The proof of the theorem depends on the following two lemmas.
Both lemmas use the contraction principle
\cite[Thm.\ VIII.3.1]{Ellis}, which states that for all $z \in [-1,1]$
\be
\label{eqn:contraction}
J_\beta (z) = \min \left\{ R(\mu | \rho_\beta): \mu \in \mathcal{P}, 
\int_\Lambda y \, \mu (dy) = z \right\}.
\ee

\begin{lemma}\per  
\label{lemma:connect} 
For $\beta > 0$ and $K > 0$
\[
\min_{\mu \in \mathcal{P}}
\left\{ R(\mu | \rho_\beta) - \beta K \left( \int_\Lambda y \, \mu
(dy) \right)^2 \right\} = \min_{|z| \leq 1} \left\{
J_\beta (z) - \beta K z^2 \right\}.
\]
\end{lemma}

\noi
{\bf Proof.} \ The contraction principle (\ref{eqn:contraction}) implies that
\begin{eqnarray*}
\lefteqn{ \min_{\mu \in \mathcal{P}}
\left\{ R(\mu | \rho_\beta) - \beta K \left( \int_\Lambda y \, \mu
(dy) \right)^2 \right\} } \\
& = & \min_{|z| \leq 1} \min \left\{ R(\mu | \rho_\beta) - \beta K
\left( \int_\Lambda
y \, \mu (dy) \right)^2 : \mu \in
\mathcal{P}, \int_\Lambda y \mu (dy) = z
\right\} \\ 
& = & \min_{|z| \leq 1} \left( \min
\left\{ R(\mu | \rho_\beta): \mu \in \mathcal{P}, \int_\Lambda y \, \mu
(dy) = z
\right\} \right) - \beta K z^2 \\ 
& = & \min_{|z| \leq 1} \left\{
J_\beta (z) - \beta K z^2 \right\}.  
\end{eqnarray*}
This completes the proof. \ \ink

\skp
The second lemma
shows that the mean of any measure $\nu \in \ebk$ is an 
element of $\tebk$.

\begin{lemma}
\label{lemma:meantovector} \per
Fix $\beta > 0$ and $K > 0$.  Given $\nu \in \mathcal{E}_{\beta,K}$,
we define $\tilde{z} \doteq \int_\Lambda y \, \nu (dy)$.  Then
$\tilde{z} \in \tilde{\mathcal{E}}_{\beta,K}$.
\end{lemma}

\noi {\bf Proof.} \ 
Since $\nu \in \mathcal{E}_{\beta,K}$, $\nu$ is a
global minimum point of $R(\mu | \rho_\beta) - \beta K (\int_\Lambda
y \, \mu (dy))^2$.  Thus for all $\mu \in \mathcal{P}$
\[
R(\nu | \rho_\beta) - \beta K \left(\int_\Lambda y \, \nu (dy)
\right)^2 = R(\nu | \rho_\beta) - \beta K \tilde{z}^2 \leq R(\mu |
\rho_\beta) - \beta K \left(\int_\Lambda y \, \mu (dy)
\right)^2.
\]
In particular, this inequality holds
for any $\mu$ that satisfies $\int_\Lambda y \, \mu (dy) =
\tilde{z}$.  For such $\mu$, the last display becomes
\[
R(\nu | \rho_\beta) \leq R(\mu | \rho_\beta).
\]
Thus $\nu$ satisfies
\[  
R(\nu | \rho_\beta) = \min \left\{ R(\mu |
\rho_\beta): \mu \in \mathcal{P}, \int_\Lambda y \, \mu (dy) = \tilde{z}
\right\}.
\]
The contraction principle (\ref{eqn:contraction}) and 
Lemma \ref{lemma:connect} imply that 
\beas 
J_\beta(\tilde{z}) - \beta K
 \tilde{z}^2 & = & R(\nu | \rho_\beta) - \beta K \left(\int_\Lambda y \,
 \nu (dy) \right)^2 \\ & = & \min_{\mu \in \mathcal{P}} \left\{R(\mu |
 \rho_\beta) - \beta K \left(\int_\Lambda y \, \mu (dy) \right)^2
 \right\} \\ & = & \min_{|z| \leq 1} \{J_\beta(z) - \beta K z^2 \}.
 \eeas 
 Therefore, $\tilde{z} \in
 \tilde{\mathcal{E}}_{\beta,K}$, as claimed.  This completes the proof. \ \ink

\skp

We next prove Theorem \ref{thm:one-to-one}.

\skp

\noi {\bf Proof of Theorem \ref{thm:one-to-one}.} \
A short calculation shows that for any $\mu \in {\cal P}$
\beas
\lefteqn{R(\mu | \rho) + \beta f_K(\mu) - \inf_{\nu \in {\cal P}}
\{R(\nu | \rho) + \beta f_K(\nu)\}} \\
&& = R(\mu | \rho_\beta) -\beta K \left(\int_\Lambda
y \, \mu(dy)\right)^2 - \inf_{\nu \in {\cal P}}
\!\left\{R(\nu | \rho_\beta) -\beta K \left(\int_\Lambda y \, \nu(dy)\right)^2\right\}.
\eeas
Hence we obtain the following alternate characterization of $\ebk$:
\be
\label{eqn:altchar}
\ebk = \left\{\nu \in \mathcal{P} : \nu 
\mbox{ minimizes } R(\nu | \rho_\beta) -\beta K \left(\int_\Lambda
y \, \nu(dy)\right)^2\right\}.
\ee

For each $z_k \in
\tilde{\mathcal{E}}_{\beta,K}$, define the set
\[
A_k \doteq \left\{\mu \in \mathcal{P} : \int_\Lambda y \, \mu
(dy) = z_k \right\}.
\]
We first show for each $k=1,..,r$, $\nu_k$ is the unique global
minimum point of $R(\mu | \rho_\beta) - \beta K \left( \int_\Lambda y \,
\mu (dy) \right)^2$ over $A_k$.  We then prove that
\[
\inf_{\mu \in A_k} \!\left\{ R(\mu | \rho_\beta) - \beta K \left(
\int_\Lambda y \, \mu (dy) \right)^2 \right\} = \inf_{\mu \in A_\ell}
\! \left\{ R(\mu | \rho_\beta) - \beta K \left( \int_\Lambda y \, \mu (dy)
\right)^2 \right\}
\]
for all $k,\ell = 1,...,r$.  It will then follow that $\{\nu_k\}_{k=1}^r$
equals the set of global minimum points of $R(\mu | \rho_\beta) -
\beta K \left( \int_\Lambda y \, \mu (dy) \right)^2$ over the set $A
\doteq \bigcup_{k=1}^r A_k$.  Finally, by showing that all the global
minimum
points of $R(\mu | \rho_\beta) - \beta K ( \int_\Lambda y \, \mu (dy))^2$
lie in $A$, we will complete the proof 
that $\ebk = \{\nu_k\}_{k=1}^r$.  If $r =2$ or 3, then since
$\int_\Lambda y \, \nu_k(dy) = z_k$, it is clear that if
$z_k \not = z_\ell$, then $\nu_k \not = \nu_\ell$.

By Theorem VIII.3.1 in \cite{EHT1}, for each $k=1,...,r$, the point $t_k$
in the statement of Theorem \ref{thm:one-to-one} exists and is unique,
\be
\label{eqn:contraction2}
J_\beta (z_k) = R(\nu_k | \rho_\beta),
\ee 
and $R(\mu|\rho_\beta)$
attains its infimum over $A_k$ at the unique measure $\nu_k$.
Therefore, for each $k=1,...,r$, $\nu_k$ is the unique global minimum
point of $R(\mu | \rho_\beta) - \beta K \left( \int_\Lambda y \, \mu (dy)
\right)^2$ over the set $A_k$.  

We next show that
\[
\inf_{\mu \in A_k} \!\left\{ R(\mu | \rho_\beta) - \beta K \left(
\int_\Lambda 
y \mu (dy) \right)^2 \right\} = \inf_{\mu \in A_\ell} \!\left\{
R(\mu | \rho_\beta) - \beta K \left( \int_\Lambda y \, \mu (dy)
\right)^2 \right\}
\]
for all $k,\ell = 1,...,r.$ Since $z_k,z_\ell \in
\tilde{\mathcal{E}}_{\beta,K}$, $z_k$ and $z_\ell$ are global minimum
points of $J_\beta (z) - \beta K z^2$.  Thus by
(\ref{eqn:contraction2}), we have 
\beas \inf_{\mu \in A_k} \! \left\{
R(\mu | \rho_\beta) - \beta K \left( \int_\Lambda y \, \mu (dy) \right)^2
\right\} & = & \inf_{\mu \in A_k} R(\mu | \rho_\beta) - \beta K z_k^2
\\ & = & J_\beta (z_k) - \beta K z_k^2 \\ & = & J_\beta (z_\ell) - \beta
K z_\ell^2 \\ & = & \inf_{\mu \in A_\ell} R(\mu | \rho_\beta) - \beta K
z_\ell^2 \\ & = & \inf_{\mu \in A_\ell} \! \left\{ R(\mu | \rho_\beta) - \beta
K \left( \int_\Lambda y \, \mu (dy) \right)^2 \right\}.  
\eeas 
As a
result, $\{\nu_k\}_{k=1}^r$ equals the set of global minimum points of $R(\mu
| \rho_\beta) - \beta K \left( \int_\Lambda y \, \mu (dy) \right)^2$ over
the set $A \doteq \bigcup_{k=1}^r A_k$.

Lastly, we show $R(\mu | \rho_\beta) - \beta K ( \int_\Lambda y \, \mu
(dy))^2$ attains its global minimum at points in $A$.  Let $\sigma$ be
a global minimum point of $R(\mu | \rho_\beta) - \beta K (
\int_\Lambda y \, \mu (dy))^2$.  By (\ref{eqn:altchar}), this implies
that
$\sigma \in \mathcal{E}_{\beta,K}$.  Define $\zeta \doteq \int_\Lambda
y \, \sigma (dy)$.  Then Lemma \ref{lemma:meantovector} implies that
$\zeta \in
\tilde{\mathcal{E}}_{\beta,K}$ and thus that $\zeta = z_k$ for some $k =
1,...,r$.  It follows that $\sigma
\in A_k \subset A$ for some $k=1,...,r$.

The last step is to prove that $t_k = 2 \beta K z_k$ for $k = 1,...,r$. From
definition (\ref{eqn:cbeta}), we have
\[
c_\beta '(t_k) = \int_\Lambda y \, \nu_k (dy) = z_k.
\]
In turn, the inverse relationship (\ref{eqn:inverse}) implies that
\[
t_k = (c_\beta ')^{-1}(z_k) = J_\beta '(z_k).
\]
Therefore, since $z_k \in \tilde{\mathcal{E}}_{\beta,K}$,
the definition 
(\ref{eqn:Snmacrostate22}) guarantees that $z_k$ is a critical point of
$J_\beta (z) -
\beta K z^2$.  Thus, 
\be
\label{eqn:ti}
t_k = J_\beta '(z_k) = 2 \beta K z_k.
\ee 
This completes the proof of Theorem \ref{thm:one-to-one}. \ \ink


\subsection{Proofs of Theorems \ref{thm:Ln2order} and
\ref{thm:Ln1order}}

Theorem \ref{thm:Ln2order} gives the structure of the set
$\ebk$ of canonical equilibrium macrostates, pointing out 
the continuous bifurcation exhibited by that set for $\beta \leq 
\beta_c \doteq \log 4$.  The structure of $\ebk$ for
$\beta > \beta_c$, given in Theorem \ref{thm:Ln1order}, features a discontinuous 
bifurcation in $K$. 
The proofs of these theorems are immediate from 
Theorems \ref{thm:secondorder} and \ref{thm:firstorder}, respectively,
which give the structure of $\tebk$ for $\beta \leq \beta_c$
and for $\beta > \beta_c$, and from Theorem \ref{thm:one-to-one}, which
states a one-to-one correspondence between 
$\tilde{\mathcal{E}}_{\beta,K}$ and $\mathcal{E}_{\beta,K}$.

Before proving Theorems \ref{thm:Ln2order} and \ref{thm:Ln1order}, 
it is useful to express the measures $\rho_\beta$
and $\nu_k$ in Theorem \ref{thm:one-to-one} in the
forms $\rho_\beta = \rho_{\beta,-1} \delta_{-1} + \rho_{\beta,0}
\delta_{0} + \rho_{\beta,1} \delta_{1}$ and $\nu_k = \nu_{k,-1}
\delta_{-1} + \nu_{k,0} \delta_{0} + \nu_{k,1} \delta_{1}$,
respectively.  Since $t_k = 2\beta K z_k$, in terms of $z_k \in \tebk$ we have
\beas
\rho_{\beta, -1} = \frac{e^{-\beta}}{1 + 2e^{-\beta}}, \hsp \rho_{\beta,
0} = \frac{1}{1 + 2e^{-\beta}}, \hsp \rho_{\beta, 1} =
\frac{e^{-\beta}}{1 + 2e^{-\beta}},
\eeas
and 
\beas
\label{eqn:nuk}
\nu_{k,-1} = \frac{e^{-2 \beta K z_k - \beta}}{C(\beta,K)}, \hsp
\nu_{k,0} = \frac{1}{C(\beta,K)}, \hsp
\nu_{k,1} = \frac{e^{2 \beta K z_k - \beta}}{C(\beta,K)}. 
\eeas
Here
\[ 
C(\beta,K) = e^{-2 \beta K z_k - \beta} + e^{2 \beta K z_k - \beta} +
1. 
\]
In particular, $\nu_k = \rho_\beta$ when $z_k = 0$.

We first indicate how Theorem \ref{thm:Ln2order} follows
 from Theorem \ref{thm:secondorder}.  Fix $\beta \leq \beta_c$.
The critical value $K_c^{(2)}(\beta)$ in Theorem \ref{thm:Ln2order}
coincides with the value $K_c^{(2)}(\beta)$ in Theorem \ref{thm:secondorder}.
For $K \leq K_c^{(2)}(\beta)$, part (a) of Theorem \ref{thm:secondorder}
indicates that 
$\tebk = \{0\}$; hence $\ebk = \{\rho_\beta\}$.
For $K > K_c^{(2)}(\beta)$, part (b) of Theorem \ref{thm:secondorder}
indicates that $\tebk = \{\pm\tilde{z}(\beta,K)\}$, where $\tilde{z}(\beta,K)
> 0$.  It follows that the measures $\nu^+$ and $\nu^-$ in part (a)(ii) of Theorem 
\ref{thm:Ln2order} are given by (\ref{eqn:macromeasure}) with $z_k = \tilde{z}(\beta,K)$
and $z_k = -\tilde{z}(\beta,K)$, respectively.
Since $\tilde{z}(\beta,K) > 0$, it follows that $\nu^+ \not = \nu^- \not = \rho_\beta$.
Finally, part (c) of Theorem \ref{thm:secondorder} allows us to 
conclude that $\nu^+$ and $\nu^-$ are continuous functions of $K >
K_c^{(2)}(\beta)$ and that both $\nu^+ \goto
\rho_\beta$ and $\nu^- \goto
\rho_\beta$ as $K \goto (K_c^{(2)}(\beta))^+$.
This completes the proof of Theorem \ref{thm:Ln2order}.

In a completely analogous way, Theorem \ref{thm:Ln1order},
including the discontinuous bifurcation noted in the last line of the theorem, follows
from Theorem \ref{thm:firstorder}.  

\skp
In this section we have completely analyzed the structure of the set 
$\ebk$ of canonical equilibrium macrostates.  In particular, we discovered 
that for $\beta \leq \beta_c$ $\, \ebk$ undergoes a continuous 
bifurcation at $K = \kconeb$ [Thm.\ \ref{thm:Ln2order}]
and that for $\beta > \beta_c$ $\, \ebk$
undergoes a discontinuous bifurcation at $K = \kctwob$ [Thm.\ \ref{thm:Ln1order}].
We depict these bifurcations in Figure 8.  
While the second-order critical values
$K_c^{(2)}(\beta)$ are explicitly defined in Theorem
\ref{thm:secondorder}, the first-order critical values
$K_c^{(1)}(\beta)$ in the figure are computed numerically.  The
numerical procedure calculates $K_c^{(1)}(\beta)$
for fixed values of $\beta$ by determining the value of $K$ for which
the number of global minimum points of $G_{\beta,K}(z)$
changes from one at $z=0$ to three at $z=0$ and $z=\pm \tilde{z}
(\beta,K)$, where $\tilde{z} (\beta,K) > 0$. According to these numerical
calculations for the discontinuous bifurcation, it appears that
$K_c^{(1)}(\beta)$ tends to $1$ as $\beta \goto \infty$.  However, we are
unable to prove this limit.

In Section 5 we will see that Figure 8 is a phase diagram that
describes the phase transitions in the canonical ensemble as
$\beta$ changes.  We will also show that the nature of the bifurcations
studied up to this point by varying $K$ while keeping $\beta$ fixed
is the same if we vary $\beta$ and keep $K$ fixed instead. The latter 
situation corresponds to what is referred to physically as a phase
transition; specifically, the continuous bifurcation corresponds to
a second-order phase transition and the discontinuous bifurcation
to a first-order phase transition.  In order to substantiate this
claim concerning the bifurcations and the phase transitions,
we have to transfer our analysis of $\ebk$ from fixed $\beta$ and
varying $K$ to an analysis of $\ebk$ for fixed $K$ and varying
$\beta$.

In the next section we study the BEG
model with respect to the microcanonical ensemble.

\begin{figure}[h]
\begin{center}
\epsfig{file=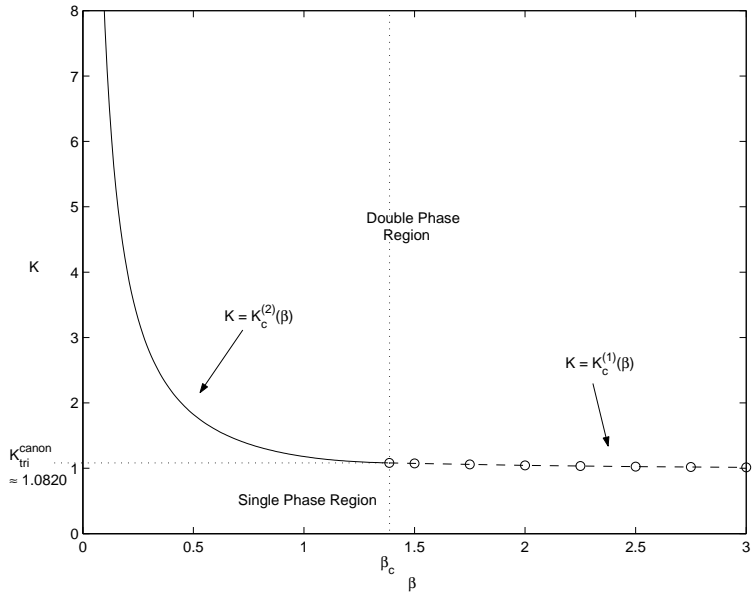,width=10cm}
\caption{ \small Bifurcation diagram for the BEG model with respect to the
canonical ensemble }
\end{center}
\end{figure}


\section{Structure of the Set of Microcanonical Equilibrium Macrostates}
\beginsec

In previous studies of the mean-field BEG model with respect to the
microcanonical ensemble, results were obtained
that either relied on a local analysis or used strictly numerical methods
\cite{BMR,BMR2,ETT}.
In this section we provide a global argument to support the existence of
a continuous bifurcation exhibited by the set
$\mathcal{E}^{u,K}$ of microcanonical equilibrium macrostates for fixed,
sufficiently large values of $u$
and for varying $K$.  Specifically, for fixed, sufficiently large $u$
$\, \euk$ exhibits a continuous bifurcation as $K$ passes through a 
critical value $K_c^{(2)}(u)$.  The
argument is similar to the one employed to analyze the canonical ensemble in
Section 3.  However, unlike the canonical case, where a rigorous
analysis of the structure of the set $\ebk$ of 
canonical equilibrium macrostates was obtained for all values of $\beta$
and $K$, the analysis of $\euk$ for sufficiently large $u$ and varying $K$
relies on a mix of analysis and numerical methods.  At the end of this
section we summarize the numerical methods used to deduce the existence
of a discontinuous bifurcation exhibited by $\euk$ for fixed, sufficiently
small $u$ and varying $K$.  In Section 5 we show how to extrapolate
this information to information concerning the phase transition behavior of the microcanonical
ensemble for varying $u$: a continuous, second-order phase transition for all 
sufficiently large values of $K$ and
a discontinuous, first-order phase transition for all sufficiently small values of $K$.

We begin by recalling several definitions from Section 2. 
$\mathcal{P}$ denotes the set of probability
measures with support $\Lambda \doteq \{-1,0,1\}$; $\rho$ denotes the measure
$\frac{1}{3}(\delta_{-1} + \delta_0 + \delta_1) \in {\cal P}$;  for
$\mu \in {\cal P}$ 
\[
R(\mu | \rho) \doteq \sum_{i=-1}^1 \mu_i \log 3\mu_i
\] 
denotes the relative entropy
of $\mu$ with respect to $\rho$; and $f_K(\mu)$ is defined by
\beas
f_K(\mu) & \doteq & \int_{\Lambda} y^2 \mu (dy) - K
\left(\int_{\Lambda} y \mu (dy) \right)^2 \\ & = & (\mu_1
+ \mu_{-1}) - K(\mu_1 - \mu_{-1})^2. 
\eeas 
For $K > 0$ we also defined the set of 
microcanonical equilibrium macrostates by 
\bea
\label{eqn:microset} 
\mathcal{E}^{u,K} & \doteq & \{ \nu
\in \mathcal{P} : I^{u,K} (\nu) = 0 \} \\ 
& = & \{ \nu \in \mathcal{P}
: \nu \ \mbox{minimizes} \ R(\nu | \rho) \ \mbox{subject to} \ f_K
(\nu) = u \},  \nonumber
\eea
$\mathcal{E}^{u,K}$ is well defined for $K > 0$ and $u \in \mbox{dom} \,
s_K = [(1-K)\wedge 0,1]$.  Throughout this section we fix $u \in \mbox{dom} \, s_K$.

Determining the elements in $\mathcal{E}^{u,K}$ requires solving a
constrained minimization problem, which is the dual of the unconstrained
minimization problem associated with the set $\mathcal{E}_{\beta,K}$ of
canonical equilibrium macrostates defined in
(\ref{eqn:canonmacrostates}).  In order to simplify the analysis of the
set $\mathcal{E}^{u,K}$, we employ the
technique used in \cite{BMR} that reduces the 
constrained minimization problem defining $\mathcal{E}^{u,K}$ to a
one-dimensional, unconstrained minimization problem.  For fixed
$K>0$ and $u \in \mbox{dom} \, s_K$, we define 
\be
\label{eqn:constraintdomain}
\mathcal{D}_{u,K} \doteq \{\mu \in \mathcal{P} : f_K(\mu) = u\}.
\ee
For
$\mu \in \mathcal{D}_{u,K}$, let $z \doteq \mu_1 - \mu_{-1}$ and $q
\doteq \mu_1 + \mu_{-1}.$  Since $\mu \in \mathcal{D}_{u,K}$ implies
that \[
f_K(\mu) \doteq (\mu_1 + \mu_{-1}) - K(\mu_1 - \mu_{-1})^2 = u,
\]
we see that $q = u + Kz^2$.  Thus, for $\mu \in
\mathcal{D}_{u,K}$, we have
\beas 
R(\mu | \rho) & = & \sum_{i=-1}^1 \mu_i \log 3 \mu_i \\
& = & \frac{q - z}{2}
\log \left[ \frac{3}{2} (q-z) \right] + (1-q) \log [ 3(1-q) ] +
\frac{q+z}{2} \log \left[
\frac{3}{2} (q+z) \right] \\ 
& = & \frac{q +
z}{2} \log (q+z) + \frac{q-z}{2} \log (q-z) \\ && \hsp + \: (1-q) \log (1-q) 
- (q \log 2 - \log 3).  
\eeas 
Setting $q = u + K z^2$, we define the quantity
\bea
\label{eqn:RuK}
R_{u,K}(z) & \doteq & \frac{q +
z}{2} \log (q+z) + \frac{q-z}{2} \log (q-z) \\
&& \hsp + \: (1-q) \log (1-q) - (q \log 2 - \log 3)
\nonumber 
\eea
and the set
\be
\label{eqn:zdomain}
\mathcal{M}_{u,K} \doteq \{z \in \R: z = \mu_1 - \mu_{-1} \ \mbox{for
some}
\ \mu \in \mathcal{D}_{u,K}\}.  
\ee
The derivation of $R_{u,K}$ makes it clear that $\mathcal{M}_{u,K} \subset
(-1,1)$ is the domain of $R_{u,K}$.  

We next introduce the set
\[
\tilde{\mathcal{E}}^{u,K} \doteq \{ \tz \in \mathcal{M}_{u,K} : \tz 
\mbox{ minimizes } R_{u,K}(z) \}.
\]
The following theorem states a one-to-one correspondence between the
elements of $\mathcal{E}^{u,K}$ and
$\tilde{\mathcal{E}}^{u,K}$.  In \cite{ETT}, for particular values of
$u$ and $K$, numerical experiments show that $\tilde{\mathcal{E}}^{u,K}$
consists of either $1,2$ or $3$ points.  Although we are not 
able to prove that this 
is valid for all $u \in \mbox{dom} \, s_K$ and $K>0$, because of
our numerical computations we make
it a hypothesis in the
next theorem.

\begin{thm} \per
\label{thm:microcorres}
Fix $K>0$ and $u \in \mbox{{\em dom}} \, s_K$.  Suppose
$\tilde{\mathcal{E}}^{u,K} = \{z_k\}_{k=1}^r$, $r = 1,..,3$.  Define
$\nu_{k}
\doteq \sum_{i=-1}^1 \nu_{k,i} \delta_i \in \mathcal{P}$ by the formulas
\[
\nu_{k,1} \doteq \frac{u + Kz_k^2 + z_k}{2}, \hsp \nu_{k,-1}
\doteq \frac{u + Kz_k^2 - z_k}{2}, \hsp \nu_{k,0} \doteq 1 -
\nu_{z_k,1} - \nu_{z_k,-1}.
\]
Then $\mathcal{E}^{u,K}$ consists of the distinct elements $\nu_{k}, k=1, \ldots, r$.
\end{thm}

\noi {\bf Proof.} Using the definition (\ref{eqn:constraintdomain}) of
$\mathcal{D}_{u,K}$, we can rewrite the set $\mathcal{E}^{u,K}$ of
microcanonical equilibrium macrostates defined in
(\ref{eqn:microset}) as
\[
\mathcal{E}^{u,K} = \{ \nu \in \mathcal{D}_{u,K} : \nu \ \mbox{is a
minimum point of} \ R(\mu | \rho) \}.
\]
We show that for $k=1,..,r$, $f_K(\nu_{k}) = u$ and $R(\nu_{k} | \rho) <
R(\mu | \rho)$ for all $\mu \in \mathcal{D}_{u,K}$ for which $\mu \neq
\nu_k$.

From the definition of $\nu_k$ we have
\[ f_K(\nu_{k}) = (\nu_{k,1} + \nu_{k,-1}) - K(\nu_{k,1} - \nu_{k,-1})^2
= (u + Kz_k^2) - Kz_k^2 = u. \] 
Therefore, $\nu_{k} \in
\mathcal{D}_{u,K}$ for all $k=1,...,r$.  
Since for all $z_k, z_\ell \in \tilde{\mathcal{E}}^{u,K}$, $k,\ell =
1,...,r$
\[
R(\nu_{k} | \rho) = R_{u,K}(z_k) = R_{u,K}(z_\ell) = R(\nu_{\ell} | \rho),
\]
it follows that $R(\nu_{k} | \rho)$ are equal for all $k=1,...,r$.

We now consider $\mu = \sum_{i=-1}^1 \mu_i \delta_i
 \in \mathcal{D}_{u,K}$ such that 
$\mu \neq \nu_{k}$ for all $k=1,...,r$.  Defining $\zeta \doteq \mu_1
- \mu_{-1}$, we claim that $\zeta \neq z_k$ for all
$k=1,...,r$.  Suppose otherwise; i.e. for some $z_k$
\be
\label{eqn:equalmean}
\mu_1 - \mu_{-1} = \zeta = z_k = \nu_{k,1} - \nu_{k,-1}. 
\ee
But $\mu \in \mathcal{D}_{u,K}$ implies that $f_K(\mu) = u = f_K(\nu_k)$
and thus that
\be
\label{eqn:equalq}
\mu_1 + \mu_{-1} = \nu_{k,1} + \nu_{k,-1}.
\ee
Combining equations (\ref{eqn:equalmean}) and (\ref{eqn:equalq}) yields
the contradiction that $\mu = \nu_k$.
Because $\zeta \neq z_k$ for all
$k=1,...,r$, it follows that $\zeta \notin \tilde{\mathcal{E}}^{u,K}$ and
thus that $R_{u,K}(z_k) < R_{u,K}(\zeta)$ for all $k=1,...,r$.  As a
result, for $k=1,...,r$ we have
\[
R(\nu_{k} | \rho) = R_{u,K}(z_k) < R_{u,K}(\zeta) = R(\mu | \rho).
\]

We complete the proof by showing that if $z_k \not = z_\ell$, then
$\nu_k \not = \nu_\ell$.  Indeed, if $\nu_k = \nu_\ell$, then
for each choice of sign we would have $Kz_k^2 \pm z_k = Kz_\ell^2 \pm z_\ell$.
Since this leads to the contradiction that $z_k = z_\ell$, the 
proof of the theorem is complete. \ \ink

\vsp Theorem \ref{thm:microcorres} allows us to analyze the set
$\mathcal{E}^{u,K}$ of
microcanonical equilibrium macrostates by
calculating the minimum points of the function $R_{u,K}$ defined in
(\ref{eqn:RuK}).  Define
\[
\varphi_{u,K} (z) \doteq \frac{q + z}{2} \log (q+z) + \frac{q-z}{2}
\log (q-z) + (1-q) \log (1-q), \mbox{ where } q = u + Kz^2.
\]
With this notation (\ref{eqn:RuK}) becomes
\[ 
R_{u,K}(z) = \varphi_{u,K}(z) - (u + Kz^2) \log 2 + \log 3.
\]

This separation of $R_{u,K}$ into the nonlinear component
$\varphi_{u,K}$ and the quadratic component is similar to the method
used in Sections 3.2 and 3.3 in determining the elements in the set
$\tilde{\mathcal{E}}_{\beta,K}$.  There we separated the minimizing
function $F_{\beta,K}(w)$ into a nonlinear component $c_\beta(w)$ and a
quadratic component $w^2/4\beta K$; minimum points of $F_{\beta,K}$
satisfy $F_{\beta,K}'(w) = c_\beta '(w) - w/2\beta K = 0$.  Solving this
equation was greatly facilitated by understanding the concavity and
convexity properties of $c_\beta$, which are proved in
Theorem \ref{thm:concavity}.

Following the success of this method in studying the canonical ensemble, we apply
a similar technique to determine the minimum points of $R_{u,K}$.  
We call a pair $(u,K)$ admissible if $u \in \mbox{dom} \, s_K$.
While an analytic proof could not be found, our numerical experiments show
that there exists a curve $K=C(u)$ in the
$(u,K)$-plane such that for all admissible $(u,K)$ lying above the graph of this curve,
$\varphi_{u,K}'$ is strictly convex on its positive domain. 
The graph of $K=C(u)$ is depicted in Figure 9.  We denote
by $G^+$ the set of admissible $(u,K)$ lying above this graph and
by $G_-$ the set of admissible $(u,K)$ lying below this graph.  Using a
similar argument as in the proof of Theorem \ref{thm:secondorder} for the canonical
case, we conclude that for all $(u,K) \in G^+$ 
the BEG model with respect to the microcanonical ensemble exhibits a
continuous bifurcation in $K$; i.e., there exists a critical
value $K_c^{(2)} (u) > 0$ such that the following hold.
\begin{enumerate}
\item
For $K \leq
K_c^{(2)}(u), \ \tilde{\mathcal{E}}^{u,K} = \{0\}.$
\item
For $K > K_c^{(2)}(u)$, there exists a positive number $\tilde{z} (u,K)$
such that $\tilde{\mathcal{E}}^{u,K} = \{\pm \tilde{z}(u,K) \}$.
\item
$\lim_{K \goto (K_c^{(2)}(u))^+} \tilde{z}(u,K) = 0$.  
\end{enumerate}

Combined with the one-to-one correspondence between the elements of
$\tilde{\mathcal{E}}^{u,K}$ and $\mathcal{E}^{u,K}$ proved in Theorem
\ref{thm:microcorres}, the structure of $\tilde{\mathcal{E}}^{u,K}$ just
given yields a continuous bifurcation in $K$ exhibited by
$\mathcal{E}^{u,K}$ for $(u,K)$ lying in the region $G^+$
above the graph of the curve
$K=C(u)$.  Similar to the definition of the critical value
$K_c^{(2)}(\beta)$ given in (\ref{eqn:2ndordercrit}) for the continuous
bifurcation in $K$ exhibited by
$\tilde{\mathcal{E}}_{\beta,K}$, the critical value $K_c^{(2)} (u)$ is
the solution of the equation
\[
R_{u,K}''(0) = 0 \hsp \mbox{or} \hsp \varphi_{u,K} ''(0) = 2K\log 2.
\]
Consequently, since $\varphi_{u,K}''(0) = 1/u + 2K[ \log (u/(1-u))]$, we
define the second-order critical value to be 
\be
\label{eqn:micro2ndcrit}
K_c^{(2)}(u) \doteq \frac{\varphi_{u,K}''(0)}{2\log 2} = \frac{1}{2u
\log \left( \frac{2(1-u)}{u} \right)}.
\ee
The derivation of this formula for $K_c^{(2)}(u)$ for the critical
values of the continuous bifurcation in $K$ exhibited by
$\mathcal{E}^{u,K}$ rests on the existence of the curve $K = C(u)$,
which in turn was derived numerically.  However the accuracy of
(\ref{eqn:micro2ndcrit}) is supported by the fact that the graph of the
curve $K_c^{(2)}(u)$ fits the critical values derived numerically in
Figures 2 and 3 of \cite{ETT}.

For values of $(u,K)$ lying in the region $G_-$ below the graph of the curve $K=C(u)$,
the strict convexity behavior of $\varphi_{u,K}'$ no longer holds.
Therefore, numerical computations were used to determine the behavior
of $R_{u,K}$ for such $(u,K)$, showing a
discontinuous bifurcation in $K$
in this region.  Specifically, there exists
a critical value $K_c^{(1)}(u)$ such that the following hold.
\begin{enumerate}
\item
For $K < K_c^{(1)}(u), \ \tilde{\mathcal{E}}^{u,K} = \{0\}$.
\item
For $K = K_c^{(1)}(u)$, there exists $\tilde{z}(u,K) > 0$
such that $\tilde{\mathcal{E}}^{u,K} = \{0, \pm \tilde{z}(u,K) \}$.
\item
For $K > K_c^{(1)}(u)$, there exists $\tilde{z}(u,K) > 0$
such that $\tilde{\mathcal{E}}^{u,K} = \{\pm \tilde{z} (u,K) \}$.  
\end{enumerate}
The critical values $K_c^{(1)}(u)$ were computed numerically by determining the
value of $K$ for which the number of global minimum points of
$R_{u,K}(z)$
changes from one at $z=0$ to three at $z=0$ and $z=\pm \tilde{z} (u,K)$,
$\tilde{z} (u,K) > 0$.  

The results of this section are summarized in
the bifurcation diagram
for the BEG model with respect to the microcanonical ensemble, which appears in Figure 9.
In the next section we will see that Figure 9 is a phase diagram that describes the 
phase transition in the microcanonical ensemble as $u$ changes.  In order to substantiate
this, we have to transfer our analysis of $\euk$ from fixed $u$
and varying $K$ to an analysis of $\euk$ for fixed $K$ and varying 
$u$.

\begin{figure}[h]
\begin{center}
\epsfig{file=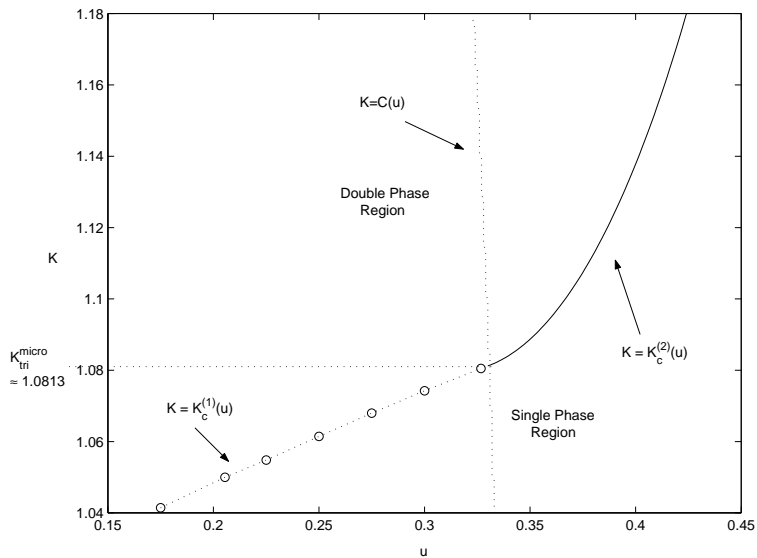,width=10cm}
\caption{ \small Bifurcation diagram for the BEG model with respect to the
microcanonical ensemble }
\end{center}
\end{figure}


\section{Comparison of Phase Diagrams for the Two Ensembles}
\beginsec

We end our analysis of the canonical and microcanonical ensembles by
explaining what our results imply concerning the nature
of the phase transitions in the BEG model.  These phase transitions are defined by
varying $\beta$ and $u$, the two parameters that define the ensembles.  As we will see,
the order of the phase transitions is a structural property of the phase diagram in
the sense that it is the same whether we vary $K$ or $\beta$ in the canonical ensemble 
and $K$ or $u$ in the microcanonical ensemble while keeping
the other parameter fixed.

Before doing this, we first review one of the main contributions of the preceding two sections,
which is to analyze the bifurcation
behavior of the sets $\ebk$ and ${\cal E}^{u,K}$ of equilibrium macrostates with respect to both the canonical and microcanonical ensembles.  
Figure 8 summarizes the canonical analysis and Figure 9 the microcanonical analysis.
The figures exhibit
two different values of $K$ called tricritical values and denoted by 
$K_{\mbox{{\scriptsize \rm tri}}}^{\mbox{{\scriptsize \rm canon}}}$ and 
$K_{\mbox{{\scriptsize \rm tri}}}^{\mbox{{\scriptsize \rm micro}}}$.
As we soon explain, at each of these values of $K$ the corresponding ensemble changes its 
behavior from a continuous, second-order phase transition to a discontinuous, first-order phase transition.
  
For the canonical ensemble, the tricritical value in Figure 8 is given by 
\[
\ktc = K_c^{(2)}(\beta_c) = K_c^{(2)}(\log 4) \approx 1.0820, 
\]
where $K_c^{(2)}(\beta)$ is defined in (\ref{eqn:2ndordercrit}).
With respect to the microcanonical
ensemble, the tricritical value $\ktm$ is the value of $K$ at which
the curves $K = C(u)$ and $K_m^{(2)}(u)$ shown in Figure 9 intersect. 
From the numerical calculation of the curve $K = C(u)$ 
we obtain the following approximation for the tricritical value $\ktm$:
\[
\ktm \approx 1.0813. 
\]
These values of $\ktc$ and $\ktm$ agree with the values derived
in \cite{BMR} via a local analysis and numerical computations.

We first illustrate how our analysis of ${\mathcal{E}}_{\beta,K}$
in Theorems \ref{thm:Ln2order} and \ref{thm:Ln1order} for fixed $\beta$
and varying $K$ yields a continuous, second-order phase transition
and a discontinuous, first-order phase transition with respect to the canonical ensemble.  
These phase transitions are defined for fixed $K$ and varying
$\beta$, the thermodynamic parameter that defines the ensemble.
In order to study the phase transition, we must therefore
transform the analysis of ${\mathcal{E}}_{\beta,K}$ 
for fixed $\beta$ and varying $K$ to an 
analysis of the same set for fixed $K$ and varying $\beta$.
After we consider the microcanonical phase transition in an analogous way, we will focus on the region
\[
\ktm \approx 1.0813 < K < 1.0820 \approx \ktc.
\]
As we will point out, 
the fact that for $K$ in this region the two ensembles exhibit different phase transition
behavior --- discontinuous for the canonical and continuous for the microcanonical --- is 
closely related to the phenomenon of ensemble nonequivalence in the model.

We begin with the continuous phase transition for the canonical ensemble.
  Figure 8 exhibits a monotonically decreasing function
$K = K_c^{(2)}(\beta)$ for $0 < \beta < \beta_c = \log 4$.  Inverting this function yields a monotonically
decreasing function $\beta = \beta_c^{(2)}(K)$ for $K > \ktc = K_c^{(2)}(\beta_c) \approx 1.0820$.
Consider for fixed $K > \ktc$ and small $\delta > 0$,
values of $\beta \in (\beta_c^{(2)}(K) - \delta, \, \beta_c^{(2)}(K) +\delta)$.  
Our analysis of ${\cal E}_{\beta,K}$ 
in Theorem \ref{thm:Ln2order} shows the following.
\begin{itemize}
\item
For $\beta \in (\beta_c^{(2)}(K) - \delta,\,\beta_c^{(2)}(K)]$
the model exhibits a single phase $\rho_\beta$.
\item For 
$\beta \in (\beta_c^{(2)}(K), \beta_c^{(2)}(K) +\delta)$ the model exhibits two distinct phases
$\nu^+(\beta,K)$ and $\nu^-(\beta,K)$.  
\end{itemize}

We claim that for fixed $K > \ktc$ 
this is a second-order phase transition; i.e., as $\beta \goto (\beta_c^{(2)}(K))^+$
we have $\nu^+(\beta,K) \goto \rho_\beta$ and $\nu^-(\beta,K) \goto \rho_\beta$.  To see this, we recall
from Figure 1(b) that for $\beta = \beta_c^{(2)}(K)$ the graph of the linear component $w/2\beta K$
of $F_{\beta,K}'(w)$ is tangent to the graph of the nonlinear component $c_{\beta}'(w)$ 
of $F_{\beta,K}'(w)$ at the origin.  This figure was used in Section 3.1 to analyze the structure
of the set $\tilde{{\cal E}}_{\beta,K}$ [Thm.\ \ref{thm:secondorder}]. Since both components of\
$F_{\beta,K}'(w)$ are continuous with respect to $\beta$, a perturbation in 
$\beta$ yields a continuous phase transition in $\tilde{\mathcal{E}}_{\beta,K}$ and thus in $\mathcal{E}_{\beta,K}$.  A similar argument shows that each of the double phases $\nu^+(\beta,K)$ and $\nu^-(\beta,K)$ are continuous functions of $\beta$ for $\beta > \beta_c^{(2)}(K)$.

We now analyze the discontinuous phase transition for the canonical ensemble in a similar way.  
Figure 8 exhibits a monotonically decreasing function
$K = K_c^{(1)}(\beta)$ for $\beta > \beta_c \doteq \log 4$.  Inverting this function yields a monotonically
decreasing function $\beta = \beta_c^{(1)}(K)$ for $K < \ktc \approx 1.0820$.
Consider for fixed $K < \ktc$ and small $\delta > 0$, values of
$\beta \in (\beta_c^{(1)}(K) - \delta, \beta_c^{(1)}(K) +\delta)$.  Our analysis of ${\cal E}_{\beta,K}$ 
in Theorem \ref{thm:Ln1order} shows the following.
\begin{itemize}
\item
For $\beta \in (\beta_c^{(1)}(K) - \delta,\beta_c^{(1)}(K))$
the model exhibits a single phase $\rho_\beta$.
\item For $\beta = \beta_c^{(1)}(K)$ the model
exhibits three distinct phases $\rho_\beta$, $\nu^+(\beta,K)$, and $\nu^-(\beta,K)$.
\item For 
$\beta \in (\beta_c^{(1)}(K), \beta_c^{(1)}(K) +\delta)$ the model exhibits two distinct phases
$\nu^+(\beta,K)$ and $\nu^-(\beta,K)$.   
\end{itemize}  

We claim that for fixed $K < \ktc$ this is a first-order phase transition; 
i.e., as $\beta \goto (\beta_c^{(1)}(K))^+$,
we have for each choice of sign 
$\nu^\pm(\beta,K) \goto \nu^\pm(\beta_c^{(1)}(K),K) \not = \rho_\beta$.  To see this, we recall
from Figure 7(a) that for $\beta = \beta_c^{(1)}(K)$ the graph of the linear component $w/2\beta K$
of $F_{\beta,K}'(w)$ intersects the graph of the nonlinear component $c_{\beta}'(w)$ 
of $F_{\beta,K}'(w)$ in five places such that the signed area between the two graphs is 0.   
This results in three values of $w$ that are global minimum points of $F_{\beta,K}$; namely,
$w =0, \tilde{w}(\beta,K), -\tilde{w}(\beta,K)$ [Thm.\ \ref{thm:firstorder}].  
These three values of $w$ give rise to three values of $z = w/2\beta K$ that
constitute the set $\tilde{{\cal E}}_{\beta,K}$ for $\beta = \beta_c^{(2)}(K)$.
Since both components of $F_{\beta,K}'(w)$ are continuous with respect to $\beta$, a perturbation in 
$\beta$ yields a discontinuous phase transition in $\tilde{\mathcal{E}}_{\beta,K}$ and thus $\mathcal{E}_{\beta,K}$.  A similar argument shows that each of the double phases $\nu^+(\beta,K)$ and $\nu^-(\beta,K)$ are continuous functions of $\beta$ for $\beta > \beta_c^{(2)}(K)$.

The phase transitions for 
the microcanonical ensemble are defined 
for fixed $K$ and varying $u$, the thermodynamic parameter defining the ensemble.
Therefore, in order to study these phase transitions, we must 
transform the analysis of ${\mathcal{E}}^{u,K}$ done in Section 4
for fixed $u$ and varying $K$ to an 
analysis of the same set for fixed $K$ and varying $u$.  This is carried out
in a way that is similar to what we have just done for the canonical ensemble.   
In particular, we find that for $K > \ktm\approx 1.0813$ the BEG model with respect to the microcanonical ensemble exhibits a continuous, second-order phase transition and that for
$K < \ktm$ the model exhibits a discontinuous, first-order phase transition.

We now focus on values of $K$ satisfying $\ktm
< K < \ktc$.  As we have just seen, for such $K$ the two ensembles exhibit different phase transition behavior: 
for $\ktm < K$ the microcanonical ensemble undergoes a continuous, second-order phase transition while
for $K < \ktc$ the canonical ensemble undergoes a discontinuous, first-order phase transition.  
This observation is consistent with a numerical calculation given in Figure 10
showing that for a fixed value of $K \in (\ktm,\ktc)$ there
exists a subset of the microcanonical equilibrium macrostates
that are not realized canonically \cite{ETT}.  As a result, for this value of $K$
the two ensembles are
nonequivalent at the level of equilibrium macrostates.

\begin{figure}[h]
\begin{center}
\epsfig{file=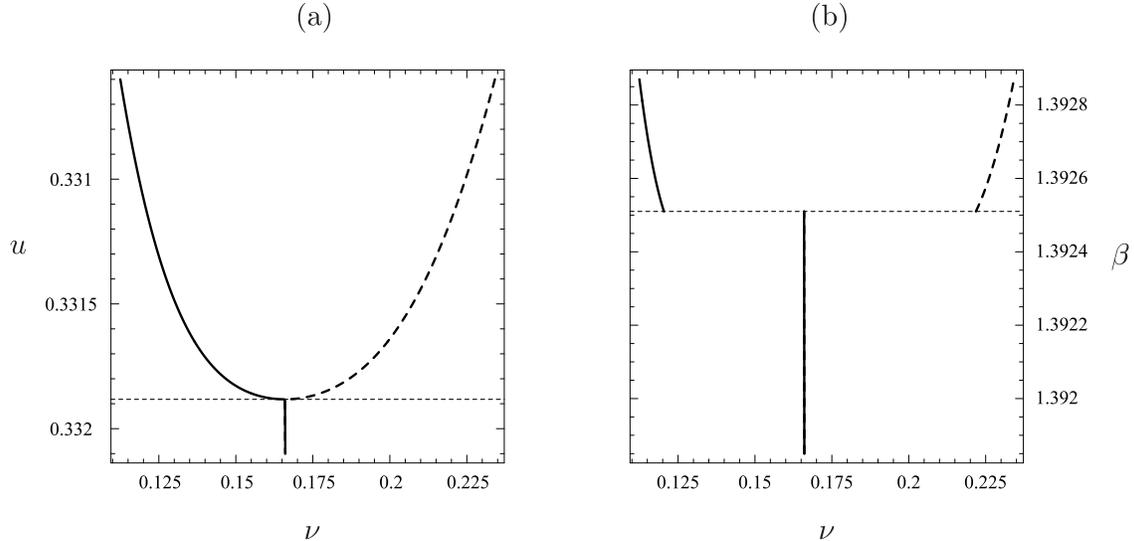,width=15cm}
\caption{ \small Structure of (a) the set ${\cal E}^{u,K}$ and (b) the set ${\cal
E}_{\beta,K}$ for $K = 1.0817$. }
\end{center}
\end{figure}

Figures 10(a) and 10(b) exhibit, for a range of values of $u$ and $\beta$, 
the structure of the set ${\cal E}^{u,K}$ of microcanonical
equilibrium macrostates and the set ${\cal E}_{\beta,K}$ 
of canonical equilibrium macrostates for $K = 1.0817$.
This value of $K$ lies in the interval $(\ktm,\ktc) \approx
(1.0813,1.0820)$.  Each equilibrium macrostate in ${\cal E}^{u,K}$ 
and ${\cal E}_{\beta,K}$ is an empirical measure having the form
\[
\nu = \nu_{1} \delta_{1} + \nu_0 \delta_0 + \nu_{-1} \delta_{-1}.
\]
In both figures the solid and dashed curves can be taken to
represent the components $\nu_{1}$ and $\nu_{-1}$.  The components
$\nu_{1}$ and $\nu_{-1}$ in the microcanonical ensemble are functions
of $u$ [Fig.\ 10(a)] and in the canonical ensemble are functions of $\beta$
[Fig.\ 10(b)].  Figures 10(a) and 10(b) were taken from \cite{ETT}.

Comparing the two figures reveals that the ensembles are nonequivalent
for this value of $K$.  Specifically, because of the discontinuous, first-order phase
transition in the canonical ensemble, there exists a subset of $\mathcal{P}$ that is not realized by
$\mathcal{E}_{\beta,K}$ for any $\beta > 0$.  On the other hand, since
the set $\mathcal{E}^{u,K}$ of microcanonical equilibrium macrostates
exhibits a continuous, second-order phase transition, the subset of
$\mathcal{P}$ not realized canonically is realized microcanonically.
As a result, there exists an inequivalence of
ensembles at the level of equilibrium macrostates.  
The reader is referred to \cite{ETT} for a more
complete analysis of ensemble equivalence and nonequivalence for the BEG model.

\section{Limit Theorems for the Total Spin with Respect to
$P_{n,\beta,K}$}
\beginsec

In Section 3.1, we rewrote the canonical
ensemble $P_{n,\beta, K}$ in terms of the total spin $S_n$ and thus reduced the
analysis of the set $\tilde{\mathcal{E}}_{\beta,K}$ of canonical
equilibrium macrostates to that of a Curie-Weiss-type model.  We end this
paper with limit theorems for the $P_{n,\beta,
K}$-distributions of appropriately scaled partial sums $S_n \doteq
\sum_{j=1}^n \omega_j$.  $S_n$ represents the total spin in the model.
Since $S_n/n = \int_{\Lambda} y \, L_n(dy)$, the limit theorems
for $S_n$ are also limit theorems for the empirical measures $L_n$.
The new theorems follow
from limit theorems for the Curie-Weiss model proved in \cite{EN} and
\cite{ENR}.

Recall the function $G_{\beta,K}$ defined in
(\ref{eqn:GbetaK}) as 
\be
\label{eqn:GbetaK2}
G_{\beta,K}(z) \doteq \beta K z^2 - c_\beta (2\beta K z), 
\ee
where
$c_\beta$ is defined in (\ref{eqn:cbeta}).  Proposition \ref{lemma:equivmin}
characterizes the set of canonical equilibrium macrostates
$\tilde{\mathcal{E}}_{\beta,K}$ by the formula
\[
\tilde{\mathcal{E}}_{\beta,K} = \{z \in \R : z \mbox{ minimizes } 
G_{\beta,K}(z) \}.
\]
In \cite{EN} and \cite{ENR}, it is proved that the limits in distribution
of appropriately scaled partial sums $S_n$ in the Curie-Weiss model are determined by the
minimum points of an analogue of $G_{\beta,K}$.  As defined in
\cite[eqn.\ (1.6)]{ENR}, this analogue is
\[
G_\beta(z) = \frac{1}{2} \beta z^2 - c_\beta(\beta z).
\]
Carrying out a similar
analysis of the minimum points of $G_{\beta,K}$ yields limit theorems
for appropriately scaled partial sums $S_n$ for the BEG model.  The
limit theorems for the BEG model are proved exactly as in the
Curie-Weiss case. 

The function $G_{\beta,K}$ is real analytic.  Hence for each global
minimum point $\tilde{z} = \tilde{z}(\beta, K)
\in \tilde{\mathcal{E}}_{\beta,K}$, there exists a positive integer
$r=r(\tilde{z})$ such that $G_{\beta,K}^{(2r)}(\tilde{z}) > 0$ and
\[
G_{\beta,K}(z) = G_{\beta,K}(\tilde{z}) +
\frac{G_{\beta,K}^{(2r)}(\tilde{z}) (z-\tilde{z})^{2r}}{(2r)!} +
O((z-\tilde{z})^{2r+1}) \ \mbox{ as } z \longrightarrow \
\tilde{z}.
\]
We call $r(\tilde{z})$ the type of the minimum point $\tilde{z}$.  This
concept is well-defined since $G_{\beta,K}$ is real analytic and
$\tilde{z}$ is a global minimum point.

Because the limiting distributions for the scaled partial sums depend on the
type of the minimum points $\tilde{z}$, we now classify each of the
points in $\tilde{\mathcal{E}}_{\beta,K}$
by type.  This is done in Theorem \ref{thm:secondordertype} 
for  $\beta \leq \beta_c$ and $K > 0$, in which case $\tebk$ exhibits
a continuous bifurcation, and in Theorem \ref{thm:firstordertype}
for $\beta > \beta_c$ and $K > 0$, in which case $\tebk$ exhibits 
a discontinuous bifurcation.  The associated limit theorems are
given in Theorems \ref{thm:uncondlimit} and \ref{thm:condlimit}.  In all cases but one 
[Thm.\ \ref{thm:secondorder}(b)] the type of each of the minimum points is 1.
When the type is 1, 
the associated limit theorems are central-limit-type theorems with
scalings $n^{1/2}$.  If $\tebk = \{0\}$, then $S_n/n^{1/2}$ converges 
in distribution to an appropriate normal random variable, and if $\tebk$ consists
of multiple points, then $S_n/n^{1/2}$ satisfies a conditioned central-limit-type 
theorem.  On the other
hand, when $K = K_c^2(\beta)$, the type of the minimum point at 0 is $r = 2$ or $r = 3$
depending on whether $\beta < \beta_c$ or $\beta = \beta_c$.  The 
associated limit theorems have non-central-limit scalings $n^{1-1/2r}$, and 
\[
P_{n,\beta,K}\{S_n/n^{1-1/2r} \in dx\} \Longrightarrow \mbox{const} \cdot
\exp[-\mbox{const}\cdot x^{2r}] \, dx.
\]
These non-classical limit theorems signal the onset of a phase transition \cite[Sect.\ V.8]{Ellis}.

We first consider $\beta
\leq \beta_c \doteq \log 4$.  According to Theorem \ref{thm:secondorder},
in this case there exists a critical value 
\be
\label{eqn:2ndordercrit2}
K_c^{(2)}(\beta) \doteq \frac{1}{2\beta c_\beta ''(0)} =
\frac{1}{4 \beta e^{-\beta}} + \frac{1}{2\beta}
\ee 
with the following properties.
\begin{itemize}
\item  For $K \leq K_c^{(2)}(\beta)$, $\tilde{\mathcal{E}}_{\beta,K}
= \{ 0 \}$.
\item For $K > K_c^{(2)}(\beta)$, there exists $\tilde{z} = \tilde{z}(\beta,K)
> 0$ such that $\tilde{\mathcal{E}}_{\beta,K} = \{ \pm \tilde{z} \}$.
\end{itemize}
The next theorem gives the type of each of these points in $\tebk$.
The type is always $1$ except when $K = K_c^{(2)}(\beta)$; in this case
the global minimum point at $0$ has type $r = 2$ if $\beta < \beta_c$
and type $r = 3$ if $\beta = \beta_c$.

\begin{thm} \per
\label{thm:secondordertype}
Let $\beta \leq \beta_c = \log 4$ and define $K_c^{(2)}(\beta)$ by
{\em(\ref{eqn:2ndordercrit2})}.  The following conclusions hold.

{\em (a)} For $K < K_c^{(2)}(\beta)$, $z=0$ has type $r=1$.

{\em (b)}  Let $K = K_c^{(2)}(\beta)$.

\hsp {\em (i)} For $\beta < \beta_c$, $z=0$ has type $r=2$.

\hsp {\em (ii)} For $\beta = \beta_c$,  $z=0$ has type $r=3$.

{\em(c)} For $K > K_c^{(2)}(\beta)$ and each choice of sign, $z=\pm \tilde{z}(\beta,K)$ has
type $r=1$.
\end{thm}

\noi {\bf Proof.} \ (a) By (\ref{eqn:2ndordercrit2}), we have
\beas
\label{eqn:gb2}
G_{\beta,K}''(0) & = & 2\beta K ( 1 - 2\beta K c_\beta ''(0)) \\ 
\nonumber & = &
2 \beta K \left( 1 - \frac{K}{K_c^{(2)}(\beta)} \right).  
\eeas
Therefore, $K < K_c^{(2)}(\beta)$ implies that $G_{\beta,K}''(0) > 0$
and thus that $z=0$ has type $r=1$.

(b) For $K = K_c^{(2)}(\beta)$, $G_{\beta,K}''(0) = 0$.  A simple
calculation yields 
\be
\label{eqn:gb4}
G_{\beta,K}^{(4)} (0) = -(2\beta K)^4 c_\beta^{(4)}(0) = -(2\beta K)^4
\cdot \frac{2e^{-\beta} ( 1 + 2e^{-\beta})(1 - 2e^{-\beta} -
8e^{-\beta})}{(1+e^{-\beta})^4}.
\ee
Therefore, for $\beta < \beta_c$, $G_{\beta,K}^{(4)}(0) > 0$ and for
$\beta = \beta_c$, $G_{\beta,K}^{(4)}(0) = 0$.  Computing the
sixth derivative yields
\be
\label{eqn:gb6}
G_{\beta_c,K}^{(6)}(0) = 2(2 \beta K)^6/9.
\ee
As a result, $z = 0$ has type 2 if $\beta < \beta_c$ and has type 3
if $\beta = \beta_c$.

(c) To prove that the symmetric minimum points $\pm \tilde{z}$ of
$G_{\beta,K}$ each have type $r=1$, we employ the results of Lemma
2.3.5 in \cite{O}.   This lemma states the
existence and uniqueness of nonzero global minimum points $\pm
\tilde{w}=\pm \tilde{w}(\beta,K)$ of $F_{\beta,K}(w) \doteq
w^2/4\beta K - c_\beta(w)$; $\tilde{w}$
is a global minimum point of $F_{\beta,K}$ if and only if $\tilde{z}
\doteq \tilde{w}/2\beta K$ is a global minimum point of $G_{\beta,K}$.
Lemma 2.3.5 in \cite{O} also states that
$F_{\beta,K}''(\tilde{w}) > 0$.  Since $F_{\beta,K}''(\tilde{w}) > 0$ if
and only if $G_{\beta,K}''(\tilde{z}) > 0$, the symmetry of
$G_{\beta,K}$ allows us to conclude that for each choice of sign $\pm \tilde{z}$ 
has type $r=1$.  This completes
the proof. \ \ink

\vsp We next classify by type the points in
$\tilde{\mathcal{E}}_{\beta,K}$
for $\beta > \beta_c$ and $K > 0$.  According to Theorem \ref{thm:firstorder}, there
exists a critical value $K_c^{(1)}(\beta)$ with the following properties.
\begin{itemize}
\item For $K < K_c^{(1)}(\beta)$, $\tilde{\mathcal{E}}_{\beta,K} =
\{ 0 \}$.
\item For $K = K_c^{(1)}(\beta)$, there exists $\tilde{z} =
\tilde{z}(\beta,K) > 0$ such that $\tilde{\mathcal{E}}_{\beta,K} = \{ 0,
\pm \tilde{z} \}$.
\item For $K > K_c^{(1)}(\beta)$, $\tilde{\mathcal{E}}_{\beta,K} = \{ \pm
\tilde{z} \}$.  
\end{itemize}
The next theorem shows that the type of each of these points in $\ebk$ is
$1$.

\begin{thm} \per
\label{thm:firstordertype}
Let $\beta > \beta_c$ and $K > 0$.  The points in 
$\tilde{\mathcal{E}}_{\beta,K}$ all have type $r=1$.
\end{thm}

\noi {\bf Proof.} \ We first consider when $\tebk$ contains 0, 
in which case $K \leq K_c^{(1)}(\beta)$.
Define $K_2 \doteq 1/2\beta c_{\beta}''(0)$.
It is proved in Lemma 2.3.15 in \cite{O} that 
for $\beta > \beta_c$ we have $K_c^{(1)}(\beta) < K_2$.
Since
\bea
\label{eqn:Gatzero}
G_{\beta,K}''(0) & = & 2\beta K ( 1 - 2\beta K c_\beta ''(0)) \\ & = &
2 \beta K \left( 1 - \frac{K}{K_2} \right), \nonumber 
\eea 
it follows that whenever
$K \leq K_c^{(1)}(\beta)$, $1 > K/K_2$ and thus $G_{\beta,K}''(0) > 0.$
We conclude that the minimum point of $G_{\beta,K}$ at
$z=0$ has type $r=1$, as claimed.

For $K \geq K_c^{(1)}(\beta)$, $\tebk$ also contains the
symmetric, nonzero minimum points $\pm \tilde{z}$ of $G_{\beta,K}$.
We prove that each of these points has type $r = 1$, employing
the results of Lemma 2.3.11 in \cite{O}.
This lemma states the existence and uniqueness of
nonzero global minimum points $\pm \tilde{w}=\pm \tilde{w}(\beta,K)$ of
the function $F_{\beta,K}(w) \doteq w^2/4\beta K - c_\beta(w)$; $\tilde{w}$ is a global minimum point of
$F_{\beta,K}$ if and only if $\tilde{z} \doteq \tilde{w}/2\beta K$ is a
global minimum point of $G_{\beta,K}$.  Furthermore, Lemma
2.3.11 states that $F_{\beta,K}''(\tilde{w}) > 0$.  Since
$F_{\beta,K}''(\tilde{w}) > 0$ if and only if $G_{\beta,K}''(\tilde{z})
> 0$, the symmetry of $G_{\beta,K}$ allows us to conclude that 
for each choice of sign $\pm \tilde{z}$ has type $r=1$.  This completes the proof. \
\ink

\vsp We now state the limit theorems for the $P_{n,\beta,
K}$-distributions for appropriately scaled partial sums $S_n$.  The
first, Theorem \ref{thm:uncondlimit}, states limit theorems that are valid
when $G_{\beta,K}$ has a unique global
minimum point at $z=0$.  This is the case for $\beta \leq \beta_c$,
$K \leq K_c^{(2)}(\beta)$ [Thm.\ \ref{thm:secondorder}(a)]
and for $\beta > \beta_c$, $K < K_c^{(1)}(\beta)$ [Thm.\ \ref{thm:firstorder}(a)].
The second, Theorem \ref{thm:condlimit}, states
conditioned limit theorems 
that are valid when $G_{\beta,K}$ has multiple global minimum points. 
Because Theorems \ref{thm:uncondlimit} and \ref{thm:condlimit}
are immediate applications of
Theorems 2.1 of \cite{EN} and 2.4 of \cite{ENR}, respectively,  we
state them here without proof.  

In Theorem \ref{thm:uncondlimit} $f_{0,\sigma^2}$ denotes the density
of an $N(0,\sigma^2)$ random variable with 
\be
\label{eqn:sigma2}
\sigma^2 \doteq \frac{2 \beta K c_{\beta}''(0)}{G_{\beta,K}''(0)}.
\ee
When the type of the minimum point at $0$ is $r = 1$, $\sigma^2 > 0$ because
in this case $G_{\beta,K}''(0) >0$ and in general $c_{\beta}''(0) > 0$.
If $f$ is a nonnegative, integrable function on $\R$, then
for $r =1$, 2, or 3 we write 
\[
P_{n,\beta,K}\{S_n/n^{1-1/2r} \in dx\} \Longrightarrow f(x) \, dx
\]
to mean that as $n \goto \infty$ the $P_{n,\beta,K}$-distributions of $S_n/n^{1-1/2r}$
converge weakly to a distribution with density proportional to $f$.  
When $r = 1$, $f = f_{0,\sigma^2}$, and the limit is
a central-limit-type theorem with scaling $n^{1/2}$.  When $r = 2$ or 3, the limits
involve the nonclassical scaling $n^{3/4}$ or $n^{5/6}$, respectively,
and the $P_{n,\beta,K}$-distributions of the scaled random variables converge weakly 
to a distribution having a density proportional
to $\exp[-\mbox{const} \cdot z^4]$ or $\exp[-\mbox{const} \cdot z^6]$.

\begin{thm} \per
\label{thm:uncondlimit}
Suppose that $\tilde{\mathcal{E}}_{\beta,K} = \{ 0 \}$ and
let $r$ be the type of the point $z=0$ as given in
Theorems {\em \ref{thm:secondordertype}} and {\em \ref{thm:firstordertype}}.  
With $\sigma^2$ the positive quantity defined
in {\em (\ref{eqn:sigma2})}, as $n \goto \infty$
\[
P_{n,\beta,K}\!\left\{\frac{S_n}{n^{1-1/2r}} \in dx\right\}
\Longrightarrow \left\{ \begin{array}{ll}
f_{0,\sigma^2}(x) \, dx & \mbox{ for } r=1 \\
\exp(-G_{\beta,K}^{(2r)}(0) x^{2r}/(2r)!) \, dx & \mbox{ for } r = 2 \mbox{ or }
r =3.
\end{array} \right. 
\] 
When $r = 2$ {\em [}$K = K_c^{(2)}(\beta)$, $\beta < \beta_c${\em ]},
$G_{\beta,K}^{(4)}(0)$ is given by {\em (\ref{eqn:gb4})}, and 
when $r = 3$ {\em [}$K = K_c^{(2)}(\beta)$, $\beta = \beta_c${\em ]},
$G_{\beta,K}^{(6)}(0) = 2(2 \beta K)^6/9$.  

\end{thm}

Our last theorem states conditioned limit theorems 
that are valid when $G_{\beta,K}$ has multiple minimum points. 
This holds in three cases: (1) when $\beta \leq \beta_c$ and 
$K > K_c^{(2)}(\beta)$, in which case the minimum points 
are $\pm \tilde{z}(\beta,K)$ with $\tilde{z}(\beta,K) > 0$
[Thm.\ \ref{thm:secondorder}(b)]; (2) when $\beta = \beta_c$
and $K = K_c^{(1)}(\beta)$, in which case the minimum points
are $0,\pm\tilde{z}(\beta,K)$ with $\tilde{z}(\beta,K) > 0$
[Thm.\ \ref{thm:firstorder}(b)]; (3) when $\beta = \beta_c$
and $K > K_c^{(1)}(\beta)$, in which case the minimum points
are $\pm \tilde{z}(\beta,K)$ with $\tilde{z}(\beta,K) > 0$
[Thm.\ \ref{thm:firstorder}(c)].
In each case in which $G_{\beta,K}$ has multiple minimum points, Theorems \ref{thm:secondordertype}and
\ref{thm:firstordertype} states that all the minimum points have type $r = 1$.
Hence, if we denote the minimum points by ${z}_j$ for $j = 1,2$ or for 
$j =1, 2, 3$, then for each $j$ we have $G_{\beta,K}''({z}_j) > 0$.
Defining
\be
\label{eqn:sigma22}
\sigma_j^2 \doteq \frac{2 \beta K c_\beta''(2 \beta K {z}_j)}{G_{\beta,K}''({z}_j)},
\ee
we see that $\sigma_j^2 > 0$.

\begin{thm} \per
\label{thm:condlimit}
Suppose that 
$\tilde{\mathcal{E}}_{\beta,K} = \{{z}_1,...,{z}_m \}$ for $m = 2$
or 3.  For each $j = 1, \ldots,m$ we 
let $f_{0,\sigma_j^2}$ be the density of an $N(0,\sigma_j^2)$ random variable, 
where $\sigma_j^2$ is the positive quantity defined in {\em (\ref{eqn:sigma22})}.
Then there exists $A = A({z}_j)
> 0$ such that for any $a \in (0,A)$
\[
P_{n,\beta,K}\!\left\{\frac{S_n - n z_j}{n^{1/2}} \in dx \left| \, 
\frac{S_n}{n} \in
[{z}_j-a,{z}_j+a] \right.\right\}
\Longrightarrow f_{0,\sigma_j}(x) \, dx \mbox{ as } n \goto \infty.
\]
\end{thm}

\skp
This completes our study of the limits for the $P_{n,\beta,
K}$-distributions of appropriately scaled partial sums $S_n \doteq
\sum_{j=1}^n \omega_j$.

\begin{acknowledgments}
We thank Marius Costeniuc for supplying the proof
of Proposition \ref{lemma:equivmin}.  The research of Richard S.\ Ellis and Peter Otto was supported by
a grant from the National Science Foundation (NSF-DMS-0202309),
and the research of Hugo Touchette was supported by
the Natural Sciences and Engineering Research Council of Canada and the Royal
Society of London (Canada-UK Millennium Fellowship).
\end{acknowledgments}

\end{document}